%% file: main.tex
\documentclass[11pt,leqno]{amsart}

\input{commands.tex}

\title[Joint discrete and continuous matrix distribution modelling]{Joint discrete and continuous matrix distribution modelling}
\date{July 2022}
\author{Martin Bladt and Clara Brimnes Gardner}
\begin{document}

\maketitle
\input{Abstract}
\input{Introduction}
\input{PH}
\input{Model}
\input{InsuranceExample}

\input{Conclusion}

\bibliographystyle{apalike}
\bibliography{main.bib}
\end{document}

%% file: commands.tex
\usepackage{amsmath,epsfig,graphicx,color, float}
\usepackage{subcaption}
\usepackage{relsize}
\usepackage{comment}
\usepackage[colorlinks, linkcolor=blue,citecolor=blue]{hyperref}
\usepackage{pdfpages}
\usepackage{graphicx}
\usepackage{mwe}
\usepackage{algorithm}
\usepackage{algorithmicx}
\usepackage{algpseudocode}
\usepackage{natbib}
\usepackage{bbm}
\usepackage{dsfont}
\DeclareSymbolFontAlphabet{\amsmathbb}{AMSb}%
\usepackage[margin=1.3in]{geometry}
\usepackage{xcolor}
\usepackage{tikz}
\newcommand{\abs}[1]{\left| #1 \right|}

\newtheorem{theorem}{Theorem}

\newtheorem{lemma}[theorem]{Lemma}
\newtheorem{example}[theorem]{Example}
\newtheorem{remark}[theorem]{Remark}

\usepackage{algorithm}
\usepackage{algpseudocode}

%% file: Abstract.tex
\begin{abstract}
In this paper we introduce a bivariate distribution on $\mathbb{R}_{+} \times \mathbb{N}$ arising from a single underlying Markov jump process. The marginal distributions are phase-type and discrete phase-type distributed, respectively, which allow for flexible behavior for modeling purposes. We show that the distribution is dense in the class of distributions on $\mathbb{R}_{+} \times \mathbb{N}$ and derive some of its main properties, all explicit in terms of matrix calculus. Furthermore, we develop an effective EM algorithm for the statistical estimation of the distribution parameters. In the last part of the paper, we apply our methodology to an insurance dataset, where we model the number of claims and the mean claim sizes of policyholders, which is seen to perform favorably. An additional consequence of the latter analysis is that the total loss size in the entire portfolio is captured substantially better than with independent phase-type models. 
\end{abstract}

%% file: Introduction.tex
\section{Introduction}

Bivariate distributions where one component is continuous and the other discrete arise naturally in many areas of application. One of the first joint continuous and discrete distributions was introduced in \cite{Bayes} in a two-step Bayesian framework: one white billiard ball is thrown onto a billiard table, the first coordinate recorded, then $n$ red billiard balls are also thrown onto the table, and the number of red balls to the left of the white ball is also recorded. 
The idea of using conditional distributions to specify mixed data distributions has since been further developed and nowadays the literature is large and growing. In the field of hierarchical models, mixed data distributions have gained notoriety (cf. \cite{GLM}.), where the continuous component is rarely observed and is introduced as an auxiliary variable handling over-dispersion. However, the Bayesian estimation of these models is then straightforward.

Another popular model to create bivariate discrete and continuous dependence is to consider fixed marginals and combine them together using a copula (cf. \cite{Nelsen}). This method has the advantage that the marginal distributions can be modeled separately from the dependence structure, but the construction has been criticized for being overly simplistic. Indeed, for almost all choices of copulas, the dependence structure cannot have arbitrarily flexible forms. The more recent data-driven approaches used in vine copulas \cite{joe2011dependence} have somewhat mitigated the latter drawback, though the simplifying assumption still poses some restrictions (see \cite{mroz2021simplifying}). In any case, these models do not have any physical interpretation and may be seen as computational tools to directly target conditional distributions.

In this paper, we introduce a flexible bivariate distribution on $\mathbb{R}_{+} \times \mathbb{N}$ through a simple and intuitive construction based on a single underlying Markov jump process. That is, we consider a joint distribution for the modeling of positive data where one marginal is continuous and the other discrete, which is both natural from a probabilistic perspective, and tractable from a statistical one. The model enjoys straightforward calculation of various quantities of interest, as well as maximum-likelihood estimation through the reconstruction of the latent (unobserved) paths of the underlying process, that is by means of the expectation-maximization algorithm. Our model is constructed with causal interpretations in mind, since the resulting observations may be thought of as rewards collected through the traversing of a Markov jump process, from inception to absorption. 

The marginals of the resulting construction are discrete phase-type distributed and (continuous) phase-type distributed, respectively. Phase-type distributions are very versatile models, and the class of continuous phase-type distributions are dense in distributions on $\mathbb{R}_{+}$. There are numerous examples of applications of continuous phase-type distributions. See for example \cite{Fackrell} for a survey of usage in the health care industry, and \cite{Mogens} for a review of the usage in risk theory. Discrete phase-type distributions have recently been used in population genetics in \cite{Hobolth} and transport modelling in \cite{Mig}. Such versatility can be adapted and extended to our setting so that our resulting model is dense on all distributions on $\mathbb{R}_{+} \times \mathbb{N}$, which makes it an ideal candidate to model data of mixed type. In particular, a dimension augmentation will be required in order to provide closed-form formulas of distributional characteristics and of the estimation algorithm.

An illustration of our model is performed on real-life insurance data. Namely, we consider the number of claims and the mean size of the claims from a Swedish motorcycle insurer. Mixed data in this setting is particularly relevant to actuarial practitioners, however, the current gold standard is to model both marginals separately and combine the two components using the independence assumption. The latter combination principle historically stems from the compound Poisson model introduced in \cite{Lundberg}, but numerous empirical studies - particularly in the industry - have shown that it is often violated in practice. For insurance purposes, some authors ( \cite{Czado,Kramer,Frees}) use the previously discussed copula approach to generate dependence, while others use latent variables (a hidden Markov model for the risk profiles in \cite{Verschuren}, and a hierarchical structure in \cite{Pinquet}). Our model also uses a latent stochastic structure to create dependence, but has the advantage that the construction has a seamless conceptual link to the development of claims, or of policyholders traversing through life stages.

The remainder of the paper is structured as follows. In Section \ref{sec:PH} we present some basic results from phase-type distribution theory, which are required in the definition of our model. In Section \ref{sec:Model} we develop our joint model, cast it in terms of reward-collecting, prove denseness, and provide distributional characteristics and statistical estimation through a dimension augmentation representation.
Subsequently, Section \ref{sec:Insurance} illustrates the flexibility and practical applicability of the model to a real-world motorcycle insurance dataset, together with some relevant discussions and associated simulated data experiments. Finally, Section \ref{sec:con} concludes.

%% file: PH.tex
\section{Preliminaries on Phase-type distributions} \label{sec:PH}
This section introduces some key results regarding Markov chains, continuous time Markov jump processes, and discrete and continuous phase-type distributions. We restrict ourselves to definitions and properties which are useful in the sequel, and refer the interested reader to the comprehensive review of these concepts found in \cite{Bo}. Throughout this section and the rest of the paper, we let $\mathbf{I}$ denote the identity matrix, $\mathbf{e}$ denote a column vector of ones, and $\mathbf{0}$ denote a matrix or a vector of zeroes, all of the appropriate dimensions.

\subsection{Markov chains and Markov jump processes}
A discrete time stochastic process $\{X_n\}_{n\in \mathbb{N}}$ on a finite state-space, $E$, is said to be a (discrete time) Markov chain if it satisfies the Markov Property
\begin{equation}
    \mathbb{P}\left(X_{n+1} = j \mid X_{n} = i_n, X_{n-1} = i_{n-1}, \ldots, X_0 = i_0 \right) = \mathbb{P}\left( X_{n+1} = j \mid X_{n} = i_n \right), \label{eq:DMarkovProperty}
\end{equation}
for $j, i_n, i_{n-1}, \ldots i_0 \in E$. If the transition probabilities in Equation \eqref{eq:DMarkovProperty} do not depend on $n$ the Markov chain is said to be time homogeneous, and we write $\mathbb{P}\left( X_{n+1} = j \mid X_{n} = i \right) = p_{ij}$. In this paper we only consider time-homogeneous Markov chains. Such a discrete time Markov chain can be completely characterized by a vector of initial probabilities $\boldsymbol{\alpha}$ and a transition matrix $\mathbf{P} = \{p_{ij}\}$. The dynamics are then described through the $n$-step transition probabilities and we have
\begin{equation*}
    \mathbb{P}\left(X_n = j \mid X_m = i\right) = \left(\mathbf{P}^{n-m}\right)_{ij},
\end{equation*}
for $i,j \in E$ and $n>m\geq 0$.

Similarly a continuous time stochastic process $\{X_t\}_{t \leq 0}$ on a finite state space, $E$, is said to be a Markov jump process (or a continuous time Markov chain) if for $t_{n+1}>t_n > t_{n-1} > \cdots t_0 > 0$ and $j, i_n, i_{n-1}, \ldots 0 \in E$ satisfies 
\begin{equation}
    \mathbb{P}\left(X_{t_{n+1}} = j \mid X_{t_{n}} = i_{n}, X_{t_{n-1}} = i_{n-1}, \ldots,  X_{t_{0}} = i_0   \right) =  \mathbb{P}\left(X_{t_{n+1}} = j \mid X_{t_{n}} = i_{n} \right). \label{eq:CMarkovProperty}
\end{equation}
If the transition probabilities in Equation \eqref{eq:CMarkovProperty} do not depend on $t$, the process is called a time-homogeneous Markov jump process. In this paper we only consider time-homogeneous Markov jump processes, although most of the results can easily be extended (with a heavier notation) to the non-homogeneous setting. 

Let $0 = t_0 < t_1 <t_2 < \cdots$ be the times of jumps in the Markov jump process. The time between jumps $S_1 = t_1 - t_0, S_2 = t_2 - t_1 \cdots$ are called the sojourn-times, and the sojourn times are exponentially distributed with a rate depending on the current state of the Markov jump process. A Markov jump process can be completely characterized by the vector of initial probabilities, $\boldsymbol{\alpha}$ and an intensity matrix, $\mathbf{S} = \{s_{ij}\}$, which represents the rates of the process. More specifically an element in the diagonal, $-s_{ii}$, $i \in E$, is the rate of the sojourn time in state $i$, and an element in the off-diagonal, $s_{ij}$, $i,j \in E, i\neq j$, specify the rate from state $i$ into state $j$. The intensity matrix thus describes the dynamics of the Markov jump process, and we have
\begin{align}
    \mathbb{P}\left(X_{s} = j \mid X_t = i\right) = \exp\left(\mathbf{S} (s-t)\right)_{ij}, \label{eq:MCDynamics}
\end{align}
for $s>t\geq 0$ and $i,j \in E$.

Every continuous time Markov jump process has a discrete time Markov chain associated with it called the embedded Markov chain. Let $\{X_t\}_{t \geq 0}$ be a Markov jump process, and let $\{Z_n\}_{n \in \mathbb{N}}$ be the embedded Markov chain. The embedded Markov chain, $\{Z_n\}_{n \in \mathbb{N}}$, records the states the Markov jump process visits,  $\{X_t\}_{t \geq 0}$, but not the sojourn times. Let $0= t_0 < t_1 <t_2 < \ldots$, be the times of the jumps in the Markov jump process. Then  $\{Z_n\}_{n \in \mathbb{N}}$ is defined by $$Z_k = X_{t_k^+},\quad k\ge0.$$ Let $\mathbf{S}$ be the intensity matrix of $\{X_t\}_{0 \leq t}$ and let $\mathbf{P}$ be the transition matrix of $\{Z_n\}_{n \in \mathbb{N}}$. The elements of $\mathbf{P}$ can the be calculated from $\mathbf{S}$ as
\begin{align}
    p_{ij} = \begin{cases} \dfrac{-s_{ij}}{s_{ii}} \quad & \text{if } i \neq j \\
    0 \quad & \text{if }i = j\end{cases}. \label{eq:EmbeddedTrans}
\end{align}
The embedded Markov chain plays a key role in the formulation of our joint continuous and discrete distribution.

\subsection{Phase-type distributions}
Both discrete time Markov chains and continuous time Markov jump processes may be defined in a way such that they have an absorbing state, which is a state that cannot be exited once it is reached. The existence of such an absorbing state leads to the concept of phase-type distributions. Consider a discrete time Markov chain $\{X_n\}_{n \in \mathbb{N}}$ with initial probability vector, $\boldsymbol{\alpha}$, transition matrix, $\mathbf{P}$, and with state space $E = \{1,2 \ldots, p, p+1\}$, where $p+1$ is an absorbing state, and the rest of the states are transient, and let $N$ denote the time until absorption
\begin{equation*}
    N = \inf\{n\geq 1 \mid X_{n} = p+1 \}.
\end{equation*}
It is then said that $N$ to follows a discrete phase-type distribution (DPH). As $p+1$ is an absorbing state we can write
\begin{equation*}
    \mathbf{P} = \begin{pmatrix}\mathbf{Q} & \mathbf{q} \\
    \mathbf{0} & 1\end{pmatrix},
\end{equation*}
with $(\mathbf{I}-\mathbf{Q})\mathbf{e} =\mathbf{q}$. The discrete phase-type distribution is then said to have representation $(\boldsymbol{\alpha},\mathbf{Q})$, and $\mathbf{Q}$ is called the sub-transition matrix. The density function and distribution function of a discrete phase-type distribution are given by
\begin{align*}
    f(n) &= \boldsymbol{\alpha} \mathbf{Q}^{n-1} \mathbf{q} \\
    F(n) &=1- \boldsymbol{\alpha} \mathbf{Q}^{n} \mathbf{e}.
\end{align*}
The Green matrix, $\mathbf{V} = \left( \mathbf{I}- \mathbf{Q}\right)^{-1}$, is important for the discrete phase-type distribution as $v_{ij}$ specifies the mean number of visits to state $j$ given that $X_0 = i$. The expected number of visits to each state can therefore be found as the elements of $\boldsymbol{\alpha} \mathbf{V}$, and the expected value of $N$ can be found by summing this vector
\begin{equation*}
    \mathbb{E}\left(N \right) = \boldsymbol{\alpha} \mathbf{V} \mathbf{e}.
\end{equation*}

Now consider a continuous time Markov jump process, $\{X_t\}_{t \geq 0}$ with initial probability vector, $\boldsymbol{\alpha}$, intensity matrix, $\mathbf{S}$, and state space $E = \{1,2 \ldots, p, p+1\}$, where $p+1$ is an absorbing state. Similarly to the discrete case we can write the intensity matrix, $\mathbf{S}$, as
\begin{align*}
    \mathbf{S} = \begin{pmatrix}\mathbf{T} & \mathbf{t} \\ \mathbf{0} & 0 \end{pmatrix},
\end{align*}
with $-\mathbf{T}\mathbf{e} = \mathbf{t}$.

We let $Y$ denote the time until absorption
\begin{equation*}
    Y = \inf\{t > 0 \mid X_t = p+1\}.
\end{equation*}
We say that $Y$ is phase-type distributed with representation $(\boldsymbol{\alpha}, \mathbf{T})$. Since we use similar calculations later we now sketch the derivation of their density function and distribution function in more detail than the discrete case. The density function and distribution function for a phase-type distribution are
\begin{align*}
    f\left(y\right) &= \boldsymbol{\alpha} \exp\left(\mathbf{T}y\right)\mathbf{t} \\
    F\left(y\right) &= 1-\boldsymbol{\alpha} \exp\left(\mathbf{T}y\right) \mathbf{e}.
\end{align*} 
The density function of $y$ can be calculated by conditioning on the initial state, $X_0$, and the state attained just before absorption, $X_{y^{-}}$. By the dynamics of the underlying Markov jump process in Equation \eqref{eq:MCDynamics} we then have
\begin{align}
\begin{split}
    f(y) &= \mathbb{P}\left(Y \in [y, y+dy]\right) \\
    &= \sum_{j=1}^{p}\sum_{i=1}^{p} \mathbb{P}\left(Y \in [y, y+dy] \mid X_0 = i, X_{y^{-1} = j}\right) \cdot \mathbb{P}\left(X_{y^{-}} = j \mid X_0 = i\right) \mathbb{P}\left(X_0 = i \right) \\
    &= \sum_{j=1}^{p} \sum_{i=1}^{p} t_j \exp\left(\mathbf{T}y\right)_{ij} \alpha_i = \boldsymbol{\alpha} \exp\left(\mathbf{T}y \right)\mathbf{t}. \end{split}\label{eq:PHDens}
\end{align}
The expression for the distribution function can then be found by integration
\begin{align*}
    F(y) = \int_{0}^y \boldsymbol{\alpha} \exp\left(\mathbf{T} u \right) \mathbf{t} du 
    = \left[\boldsymbol{\alpha} \exp\left(\mathbf{T} u \right) \mathbf{T}^{-1} \mathbf{t}\right]_0^y 
    = 1- \boldsymbol{\alpha} \exp\left(-\mathbf{T} y \right) \mathbf{e}.
\end{align*}

As for the discrete phase-type distribution, the Green matrix, $\mathbf{U} = (-\mathbf{T})^{-1}$, has elements $u_{ij}$ that specifies the mean time spend in each state given the initial state. The elements of $\boldsymbol{\alpha} \mathbf{U}$ specifies the expected time spend in each state, and the expected value of $Y$ is obtained by summing up all the elements of the vector:
\begin{equation*}
    \mathbb{E}\left(Y\right) = \boldsymbol{\alpha} \mathbf{U}\boldsymbol{e}
\end{equation*}

%% file: Model.tex
\section{Joint discrete and continuous Phase-type distributions} \label{sec:Model}
In this section we present the main model for joint bivariate and continuous and discrete phase-type distributions. As for the univariate case, we let $\{X_t\}_{0\leq t}$ be a continuous time Markov jump process on a state space $E  = \{1,2,\ldots, p, p+1\}$. The first $p$ states are transient and state $p+1$ is absorbing. Let $\boldsymbol{\alpha}$ be the initial distribution on the transient state, and let $\mathbf{T}$ be the sub-generator matrix on the transient states. Let $Y$ be the time until absorption in the Markov jump process
\begin{equation*}
    Y = \inf\{t>0 \mid X_t = p+1 \}.
\end{equation*}
Then $Y$ follows a phase-type distribution with representation $\left(\boldsymbol{\alpha}, \mathbf{T}\right)$. Let $\{Z_n\}_{n \in \mathbb{N}}$ denote the embedded Markov chain of $\{X_t\}_{0\leq t}$. We use this embedded Markov chain to construct a discrete random variable, $N$, which due to the relationship between $\{X_t\}_{0\leq t}$ and $\{Z_n\}_{n \in \mathbb{N}}$ will inherit a natural dependence structure with respect to $Y$. The obvious choice is to let $N$ be the time until absorption in $\{Z_n\}_{n \in \mathbb{N}}$. This, however, creates a natural, hard-wired positive dependence between $Y$ and $N$, which makes it inflexible and thus problematic to model, for instance, negative dependence. To circumvent this initial drawback, we construct a slight modification which turns our to be much more flexible.

Let $\tilde{E} = \{1,2,\ldots, p\}$ be the transient state space of $\{X_{t}\}_{t \geq 0}$. We partition $\tilde{E}$ into two disjoint sets, $E^{+}$ and $E^{0}$, such that $E^{+} \cup E^{0} = \tilde{E}$. We then define $N$ as the number of times $\{X_t\}_{0\leq t}$ visits states in $E^{+}$ until absorption.
\begin{equation*}
    N = \inf\left\{\sum_{i=1}^n  \mathbf{1}_{Z_i \in E^+} \mid Z_n = p+1\right\}.
\end{equation*}
Figure \ref{fig:my_label} illustrates one realization of the Markov jump process and the development of $Y$ and $N$ according to the underlying process.

\usetikzlibrary{arrows.meta} 
\begin{figure}[!htbp]
    \centering
    \begin{tikzpicture}
    \draw[-stealth,black,thick] (0,0) -- (10,0);
    \draw[-stealth,black,thick] (0,0) -- (0,4.5);
    \draw[black,dotted]   (0,0.8) -- (10,0.8);
    \draw (0,0.8) node[anchor=east]{1};
    \draw[black,dotted] (0,1.6) -- (10,1.6);
    \draw (0,1.6) node[anchor=east]{2};
    \draw[black,dotted] (0,2.4) --(10,2.4);
    \draw (0,2.4) node[anchor=east]{3};
    \draw[black,dotted] (0,3.2) --(10, 3.2);
    \draw (0,3.2) node[anchor=east]{4};
    \draw[black,dotted] (0,4) -- (10,4);
    \draw (0,4) node[anchor=east]{5};
    \filldraw [blue] (0,.8) circle (1pt);
    \draw[blue,thick] (0,.8) -- (1.155,.8);
    \draw[blue] (1.2,.8) circle (1pt);
    \filldraw[blue] (1.2,3.2) circle (1pt);
    \draw[blue, thick] (1.2,3.2) -- (1.455,3.2);
    \draw[blue] (1.5,3.2) circle (1pt);
    \filldraw[blue] (1.5,1.6) circle (1pt);
    \draw[blue,thick] (1.5,1.6) -- (3.455,1.6);
    \draw[blue] (3.5,1.6) circle (1pt);
    \filldraw[blue] (3.5,0.8) circle (1pt);
    \draw[blue,thick] (3.5,0.8) -- (4.455,0.8);
    \draw[blue] (4.5,0.8) circle (1pt);
    \filldraw[blue] (4.5,3.2) circle (1pt);
    \draw[blue,thick] (4.5,3.2) -- (4.855,3.2);
    \draw[blue] (4.9,3.2) circle (1pt);
    \filldraw[blue] (4.9,2.4) circle (1pt);
    \draw[blue,thick] (4.9,2.4) -- (5.855,2.4);
    \draw (5.9,-0.1) node[anchor = north]{$\tau$};
    \draw[black] (5.9,-0.1) -- (5.9,0.1);
    \draw[blue] (5.9,2.4) circle (1pt); 
    \filldraw[blue] (5.9,4) circle (1pt);
    \draw[blue,thick] (5.9,4) -- (10,4);
    \draw[-stealth,teal,thick] (0,-7.5) --(0,-1);
    \draw[teal,thick] (0,-7.5) -- (5.9,-1.6);
    \draw[teal,thick] (5.9,-1.6)-- (10,-1.6);
    \draw[black,thick] (0,-7.5) --(10,-7.5);
    \filldraw[violet] (0,-6.5) circle (1pt);
    \draw[violet,thick] (0,-6.5) -- (1.455,-6.5);
    \draw[violet] (1.5,-6.5) circle (1pt);
    \filldraw[violet] (1.5,-5.5) circle (1pt);
    \draw[violet] (1.5,-5.5) -- (3.455,-5.5);
    \draw[violet] (3.455,-5.5) circle (1pt);
    \filldraw[violet] (3.5,-4.5) circle (1pt);
    \draw[violet] (3.5,-4.5) -- (10,-4.5);
    \draw[-stealth,violet,thick] (10,-7.5) -- (10,-1);
    \draw[violet] (9.9,-6.5) -- (10.1,-6.5);
    \draw[violet] (9.9,-5.5) -- (10.1,-5.5);
    \draw[violet] (9.9,-4.5) -- (10.1,-4.5);
    \draw[violet] (9.9,-3.5) -- (10.1,-3.5);
    \draw[violet] (9.9,-2.5) -- (10.1,-2.5);
    \draw[violet] (9.9,-1.5) -- (10.1,-1.5);
    \draw[teal] (-0.1,-1.6) -- (0.1,-1.6); 
    \draw[black,dotted] (0,-6.5) -- (10.,-6.5);
    \draw[black, dotted] (0,-5.5) -- (10,-5.5);
    \draw[black,dotted] (0,-4.5) -- (10,-4.5);
    \draw[black,dotted] (0,-3.5) -- (10,-3.5);
    \draw[black,dotted] (0,-2.5) -- (10,-2.5);
    \draw[black,dotted] (0,-1.5) -- (10,-1.5);
    \draw[teal] (-0.1,-1.6) -- (0.1,-1.6); 
    \draw (-0.1,-1.6) node[anchor = east]{{\color{teal}$\tau$}};
    \draw (10.1,-6.5) node[anchor = west]{{\color{violet}$1$}};
    \draw (10.1,-5.5) node[anchor = west]{{\color{violet}$2$}};
    \draw (10.1,-4.5) node[anchor = west]{{\color{violet}$3$}};
    \draw (10.1,-3.5) node[anchor = west]{{\color{violet}$4$}};
    \draw (10.1,-2.5) node[anchor = west]{{\color{violet}$5$}};
    \draw (10.1,-1.5) node[anchor = west]{{\color{violet}$6$}};
    \draw (10.5,-4) node[anchor = west]{{\color{violet}$N(t)$}};
    \draw (-0.5,-4) node[anchor = east]{{\color{teal}$Y(t)$}};
    \draw[black] (5.9,-7.4) -- (5.9,-7.6);
    \draw (5.9,-7.6) node[anchor = north]{$\tau$};
    \end{tikzpicture}
    \caption{A realization of a Markov jump process and the corresponding development of $(Y,N)$. The top plot shows a path of a Markov jump process on $E = \{1,2,3,4,5\}$ with $\{1,2,3,4\}$ being transient states and $5$ being the absorbing state. Furthermore we have $E^{+} = \{1,2\}$. The process jumps between the four transient states until it at time $\tau$ is absorbed in state $5$.  Let $Y(t)$ be the value of $Y$ at time $t$ and let $N(t)$ be the value of $N$ at time $t$. The green function corresponds to $Y(t)$, while the red function shows the value of $N(t)$. Note that the x-axis on the bottom plot is the same as the x-axis on the top plot. At the time-points the Markov jump process jumps into a state in $E^{+} =\{1,2\}$ the value of $N(t)$ increases by $1$. Notice that the increment happens upon entering the states of $E^{+}$, but one could also let the increments happen upon exiting these states. This realization corresponds to to $(Y,N)=(\tau,3)$.}
    \label{fig:my_label}
\end{figure}
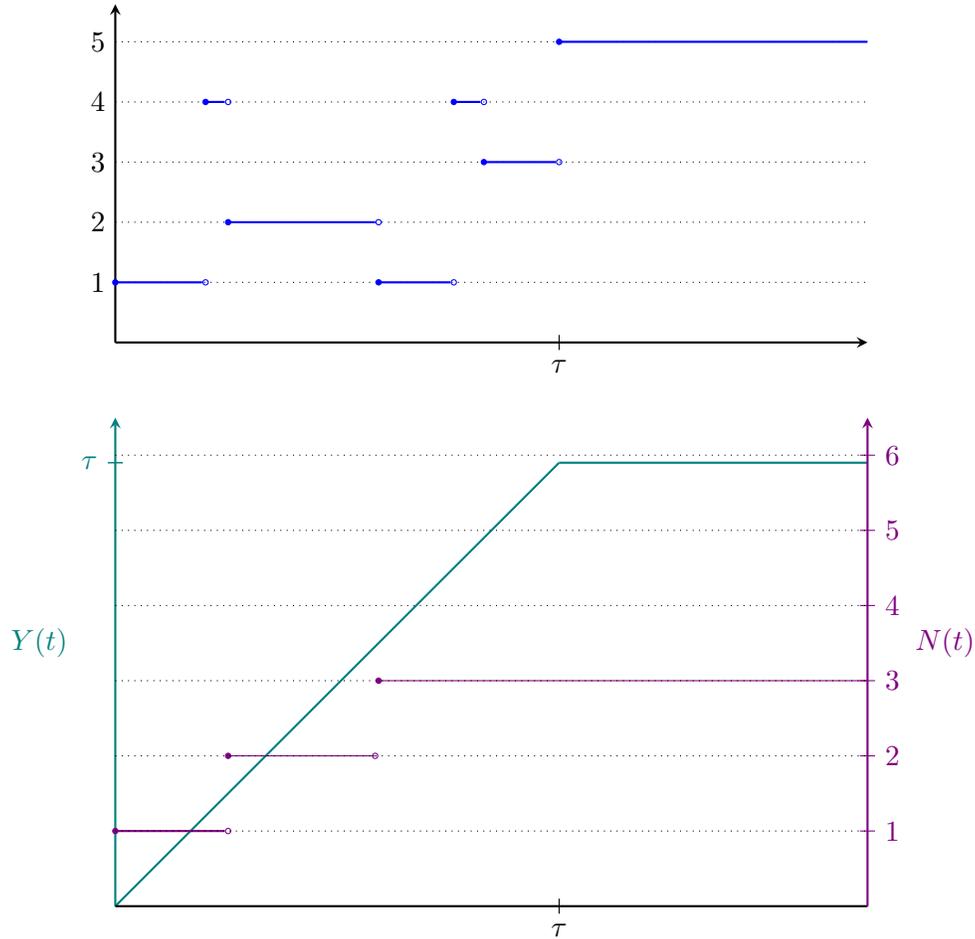


For convenience, we organize $\boldsymbol{\alpha}$ and $\mathbf{Q}$ according to the division into $E^+$ and $E^0$ such that the first states are the the states in $E^+$ and the last states are the states in $E^0$. This can be done by a simple re-labelling of states such that
\begin{align*}
    \boldsymbol{\alpha} &=\begin{pmatrix} \boldsymbol{\alpha}^+ & \boldsymbol{\alpha}^0 \end{pmatrix} \\
    \mathbf{Q} &= \begin{pmatrix} \mathbf{Q}^{++} & \mathbf{Q}^{+0} \\
    \mathbf{Q}^{0+} & \mathbf{Q}^{00}\end{pmatrix}.
\end{align*}
It can then be shown, by using transformation of rewards, that $N$ follows a DPH-distribution with representation $(\tilde{\boldsymbol{\alpha}}, \tilde{\mathbf{Q}})$, where
\begin{align}
\begin{split}
    \tilde{\boldsymbol{\alpha}} &= \boldsymbol{\alpha}^+ +   \boldsymbol{\alpha}^0 (\mathbf{I}-\mathbf{Q}^{00})^{-1} \mathbf{Q}^{+0} \\
    \tilde{\mathbf{Q}} &= \mathbf{Q}^{++} + \mathbf{Q}^{+0}(\mathbf{I}-\mathbf{Q}^{00})^{-1}\mathbf{Q}^{+0}, \end{split} \label{eq:TRV}
\end{align}
and $\boldsymbol{\alpha}^{+}$ are the probabilities of the Markov jump process, $\{X_t\}_{t\geq 0}$ starting in the states in $E^{+}$, while $\boldsymbol{\alpha}^0$ are the probabilities of  $\{X_t\}_{t\geq 0}$ staring in the states in $E^{0}$. The matrix $\mathbf{Q}^{++}$ contains the transition probabilities between states in $E^{+}$, $\mathbf{Q}^{+0}$ the transition probabilities from states in $E^{+}$ to states in $E^{0}$, and so on. We let $\{\tilde{Z}_{n}\}_{n \in \mathbb{N}}$ denote the discrete Markov chain behind the DPH-distribution with representation $(\tilde{\boldsymbol{\alpha}}, \tilde{\mathbf{Q}})$. Then
\begin{equation*}
    N = \inf\left\{\sum_{i=1}^n  \mathbf{1}_{Z_i \in E^+} \mid Z_n = p+1\right\} = \inf\{n \geq 0 \mid \tilde{Z}_n = p+1\}
\end{equation*}
The joint distribution of $(Y,N)$ is thus bivariate with phase-type and discrete phase-type marginals. The distribution is specified by the initial probabilities, $\boldsymbol{\alpha}$, the subintensity matrix, $\boldsymbol{T}$, and $E^{+}$. For the rest of this paper we assume that $\boldsymbol{\alpha}^{0} = \mathbf{0}$ and $\boldsymbol{\alpha}^{+} \mathbf{e} = 1$, as this ensures that $N \geq 1$.

Our proposed distribution may be seamlessly cast into a framework akin to MPH$^*$ distributions and MDPH$^*$ distributions. For its conceptual significance, we now highlight the connection. MPH$^*$ distributions were introduced in \cite{MPH} as a way of extending continuous phase-type distributions to the multivariate case and have stood as one of the most flexible ways of generating dependence. Consider a Markov jump process, $\{X_t\}_{t \geq 0}$ with sub-intensity matrix $\mathbf{T}$ and initial distribution $\boldsymbol{\alpha}$. Let $C_i$ be the total time spent in state $i$ until absorption, let $C_{ik}$ be the time spend in state $i$ upon the $k$th visit to the state, and let $N_i$ be the number of visits to state $i$ before absorption. Then
\begin{equation}
    C_i = \sum_{k=1}^{N_i} C_{ik} = \int_{0}^{Y} \mathbf{1}_{X_t = i} dt,
\end{equation}
where $Y$ is the time until absorption. Thus $Y \sim \text{PH}\left(\boldsymbol{\alpha}, \mathbf{T}\right)$. Furthermore,   $Y = \sum_{i=1}^{p} C_i.$
    
To create $m$ dependent PH-distributed variables all based on the same process $\{X_t\}_{t \geq 0}$, we introduce a reward matrix, $\mathbf{R} = \{r_{ij}\}$, of dimension $p \times m$, where $p$ is the number of transient states. We require $r_{ij}\geq 0$, and define
\begin{equation*}
    Y_j = \sum_{i=1}^{p} r_{ij} C_i, \quad j \in \{1,\ldots, m\}.
\end{equation*}
The r-th column of $\mathbf{R}$ can thus be interpreted as a reward-function for $Y_j$, and $Y_j$ is the accumulated reward until absorption. We write $\mathbf{Y} = \begin{pmatrix} Y_1 & Y_2 & \cdots  & Y_m\end{pmatrix} \sim \text{MPH}^*\left(\boldsymbol{\alpha}, \mathbf{T}, \mathbf{R}\right)$. A similar construction can be made for discrete random-variables, in this way giving rise to $\text{DMPH}^*\left(\boldsymbol{\alpha}, \mathbf{Q}, \mathbf{R}\right)$ distributions; we omit the details for brevity. 


Coming back to our model, we defined $Y$ as the time until absorption in the Markov jump process, and $N$ as the number of times the embedded Markov chain, $\{Z_n\}_{n \in \mathbb{N}}$, visited $E^{+}$. We can write $(Y,N)$ as a mixed MPH$^*$ and DMPH$^*$ distribution, which we will denote MMPH$^*$ for Mixed MPH$^*$. Let $\mathbf{R}$ be a $2 \times p$ reward matrix with elements
\begin{align*}
    r_{i1} & = 1 \\
    r_{i2} &= \begin{cases} 1 \quad & \text{if } i \in E^{+}  \\
    0 \quad & \text{if } i \in E^{0}\end{cases}.
\end{align*}
The first column is a continuous reward function that assigns rewards based on the time spent in each state, and the second column is a discrete reward function that assigns rewards upon entering each state. Then we can then write $(Y,N) \sim \text{MMPH}^*\left(\boldsymbol{\alpha}, \mathbf{T}, \mathbf{R} \right)$. 

While this connection to MPH$^*$ and DMPH$^*$ distributions is not necessary for the understanding of the model, it is beneficial to utilize in some of the proofs that follow in later sections. The DMPH$^*$ class is interesting in its own right, however, we presently concentrate on joint modeling of discrete and continuous distributions. For an account of the properties of the DMPH$^\ast$ class, we refer to \cite{Azucena}.

The rest of this section is devoted to properties of the MMPH$^\ast$ distribution. In Section \ref{sec:Denseness} we show that the distribution class in dense is the general class of distributions on $\mathbb{R}_{+} \times \mathbb{N}$. In Section \ref{sec:Augmenting} we present a method of augmenting the state space such that the absorption state reflects the value of $N$, and in Sections \ref{sec:JointDist}, \ref{sec:ConditionalDist} and \ref{sec:MGF} this augmented representation is used to calculate the joint distribution, the conditional distribution and the joint moment generating function, respectively. Finally, in Section \ref{sec:Estimation} we consider statistical inference via an EM algorithm.

\subsection{Denseness of model class} \label{sec:Denseness}
It is a well known fact that phase-type distributions are dense in the class of distributions on $\mathbb{R}_+$. The first proof of this fact is due to \cite{Schassberger}. Here, we extend the proof given in \cite{JohnsonTaaffe} to our setting. To this end, let $F$ be an arbitrary distribution on $\mathbb{R}_{+} \times \mathbb{N}$. For fixed $n$, consider the conditional distribution $F_{Y\mid N = n}$. This is a distribution on $\mathbb{R}_{+}$ and there exists a sequence of phase-type distributions that converges weakly to $F_{Y \mid N = n}$. We now seek to create such a phase-type representation that satisfies the additional constraint that $N = n$ is always possible for each of the approximating terms. We define three auxiliary distributions
\begin{align*}
    F^{(l)}_{Y \mid N = n}(y) &= \sum_{k=1}^{\infty} d_{F_{Y \mid N = n}}(l,k) E_k(y;l) \\
    \tilde{F}^{(l)}_{Y \mid N = n}(y) &= \sum_{k=1}^{\infty} d_{F_{Y \mid N = n}}(l,k) E_{\max{k,n}}(y;l) \\
    \tilde{F}^{(l,m)}_{Y \mid N = n}(y) &= F_{Y \mid N = n}\left(\dfrac{m}{l}\right)^{-1}\sum_{k=1}^m d_{F_{Y \mid N = n}}(l,k) E_{\max\{k,n\}}(y;l),
\end{align*}
with
\begin{align*}
    d_{F_{Y \mid N=n}}(l,k) &= F\left( \dfrac{k}{l}\right) - F\left(\dfrac{k-1}{l} \right)\\
    E_k(y;l) &= 1- \sum_{i=0}^{k-1} \dfrac{\exp\left(-ly\right) (ly)^{i}}{i!}.
\end{align*}
From \cite{JohnsonTaaffe}, we know that $F^{(l)}_{Y\mid N = n} \to F_{Y \mid N = n}$ for $l \to \infty$, which can be used to to provide the following result.

\begin{lemma}\label{lem:cond}
As $m\to\infty$ and then $l\to\infty$ we have
$$\tilde{F}^{(l,m)}_{Y \mid N = n}(y) \to F_{Y\mid N = n}(y),\quad y\ge0.$$
\end{lemma}

\begin{proof}
We show that $ \tilde{F}^{(l)}_{Y \mid N = n}(y) \to  F^{(l)}_{Y \mid N = n}(y)$ for $l \to \infty$ and that $\tilde{F}^{(l,m)}_{Y \mid N = n}(y) \to \tilde{F}^{(l)}_{Y \mid N = n}(y)$ for $m \to \infty$, which can be combined to show $\tilde{F}^{(l,m)}_{Y \mid N = n}(y) \to F_{Y\mid N = n} $ for $l,m \to \infty$. 
\begin{align*}
    \abs{\tilde{F}^{(l)}_{Y \mid N = n} - F^{(l)}_{Y \mid N = n}} &= \abs{ \sum_{k=1}^{\infty} d_{F_{Y \mid N = n}}(l,k) E_{\max(k,n)}(y;l) - \sum_{k=1}^{\infty} d_{F_{Y \mid N = n}}(l,k) E_k(y;l)} \\
    &= \abs{\sum_{k=1}^{n-1} d_{F_{Y\mid N = n}}(l,k)( E_{\max(k,n)}(y;l) - E_k(y;l))} \\
    &\leq \abs{\sum_{k=1}^{n-1} \left(1-\sum_{i=1}^{k-1} \exp\left(-ly\right) \dfrac{(ly)^{i}}{i!} \right) - \left(1-\sum_{i=1}^{n-1} \exp\left(-ly\right) \dfrac{(ly)^{i}}{i!} \right)} \\
    &= \sum_{k=1}^{n-1} \sum_{i=k}^{n-1} \exp\left(-ly\right) \dfrac{(ly)^{i}}{i!},
\end{align*}
which converges to zero for $l \to \infty$. For the next limit we can always take $m \leq n$, and so
\begin{align*}
    &\abs{\tilde{F}^{(l)}_{Y\mid N = n}(y) - \tilde{F}^{(m,l)}_{Y\mid N = n}(y)} \\ &=\abs{\sum_{k=1}^{\infty}d_{F_{Y \mid N = n}}(l,k) E_{\max(k,n)}(y;l) - F\left(\dfrac{m}{l}\right)^{-1} \sum_{k=1}^m d_{F_{Y \mid N = n}}(l,k) E_{\max(k,n)}(y;l) }\\
    &= \abs{\left(1-F_{Y\mid N = n}\left(\dfrac{m}{l} \right)^{-1} \right)\sum_{k=1}^m d_{F_{Y\mid N = n}}(l,k)  E_{\max(k,n)}(y;l) +  \sum_{k=m+1}^{\infty} d_{F_{Y \mid N = n}}(l,k) E_k(y;l)} \\
    &\leq  \abs{\left(1-F_{Y\mid N = n}\left(\dfrac{m}{l} \right)^{-1} \right)\sum_{k=1}^m d_{F_{Y\mid N = n}}(l,k) } + \abs{ \sum_{k=m+1}^{\infty} d_{F_{Y \mid N = n}}(l,k) } \\
    &\leq \abs{\left(1-F_{Y\mid N = n}\left(\dfrac{m}{l}\right)^{-1} \right) F_{Y \mid N =n}\left(\dfrac{m}{l}\right)} + 1-F_{Y\mid N = n}\left(\dfrac{m}{n}\right),
\end{align*}
and $\abs{\left(1-F_{Y\mid N = n}\left(\dfrac{m}{l}\right)^{-1} \right) F_{Y \mid N =n}\left(\dfrac{m}{l}\right)} + 1-F_{Y\mid N = n}\left(\dfrac{m}{n}\right) \to 0 $ for $m \to \infty$. Thus, putting the pieces together we obtain that
\begin{align*}
    &\abs{F_{Y\mid N = n}(y)- \tilde{F}_{Y\mid N = n}^{(m,l)}(y)} \\&= \abs{F_{Y\mid N = n}(y) - F^{(l)}_{Y\mid N = n}(y) + F^{(l)}_{Y\mid N = n}(y) - \tilde{F}^{(l)}_{Y\mid N = n}(y) + \tilde{F}^{(l)}_{Y\mid N = n}(y) - \tilde{F}_{Y\mid N = n}^{(m,l)}(y)} \\
    & \leq \abs{F_{Y\mid N = n}(y) - F^{(l)}_{Y\mid N = n}(y)}+ \abs{  F^{(l)}_{Y\mid N = n}(y) - \tilde{F}^{(l)}_{Y\mid N = n}(y)} +  \abs{\tilde{F}^{(l)}_{Y\mid N = n}(y) - \tilde{F}_{Y\mid N = n}^{(m,l)}(y)},
\end{align*}
which by our previous calculations and the result from \cite{JohnsonTaaffe} goes to $0$, for $m\to\infty$ and then $l\to\infty$. 
\end{proof}

We note that we can create a phase-type distribution with distribution function $\tilde{F}^{(m,l)}_{Y\mid N = n}$ where at least $n$ states are visited, as $\tilde{F}^{(m,l)}_{Y\mid N = n}$ is the distribution function of a mixture of $k$ Erlang distributions with at least $n$ stages. This means that by construction we can ensure $N=n$ in the phase-type distribution by letting the first $n$ states in each Erlang distribution be in $\abs{E}^{+}$. A PH representation, $(\boldsymbol{\alpha}_k, \mathbf{T}_k)$, for a $k$ stage Erlang distribution with rate $\lambda$ and  with $k$ states is
\begin{align*}
    \boldsymbol{\alpha}_k &  = \begin{pmatrix}1 & 0 & 0 & \cdots & 0 \end{pmatrix} \\
    \mathbf{T}_k(\lambda) &= \begin{pmatrix} -\lambda & \lambda & 0 & \cdots & 0 \\
    0 & -\lambda & \lambda & \cdots & 0 \\ 
    \vdots & \vdots & \vdots & \ddots & \vdots \\
    0 & 0 & 0 & \cdots & -\lambda
    \end{pmatrix}.
\end{align*}
Using this we can create a phase-type representation, $(\boldsymbol{\beta}_{n,m}(l),\mathbf{S}_{n,m}(l))$ for $\tilde{F}^{(m,l)}_{Y \mid N = n}(y)$ using $m$ of these Erlang blocks:
\begin{align}
\begin{split}
    \boldsymbol{\beta}_{n,m}(l) & = \begin{pmatrix}\frac{d_{F_{Y \mid N = n}}(l,n)}{F_{Y \mid N = n}\left(\dfrac{m}{l} \right) } & \frac{ d_{F_{Y \mid N = n}}(l,n)}{F_{Y \mid N = n}\left(\dfrac{m}{l} \right)} & \cdots \frac{ d_{F_{Y \mid N = n}}(l,m)}{F_{Y \mid N = n}\left(\dfrac{m}{l} \right)} \end{pmatrix} \\
    \mathbf{S}_{n,m}(l) & = \begin{pmatrix} \mathbf{T}_n(l) & 0 & 0 & \cdots & 0 \\
    0 & \mathbf{T}_n(l) & 0 & \cdots & 0 \\
    0 & 0 & \mathbf{T}_n(l) & \cdots & 0 \\
    \vdots & \vdots & \vdots & \ddots & \vdots \\
    0 & 0 & 0 & \cdots & \mathbf{T}_m(l)\end{pmatrix}. \end{split} \label{eq:DensenessPHCondRep}
\end{align}
The next step is to combine the conditional distributions such that we get the right joint distributional convergence. 
\begin{theorem}
For any mixed distribution function $F$ there exists a sequence of MMPH$^\ast$ distributions converging weakly to it.
\end{theorem}
\begin{proof}
We consider a generator matrix consisting of $b$ blocks. The blocks do not communicate, and the sub-generator matrix for block number $i$ is the phase-type representation of $\tilde{F}^{(m,l)}_{Y\mid N = i}(y)$. The probability of entering block $i$ is set to 
\begin{equation*}
    \tilde{p}^{(b)}_i = \dfrac{\mathbb{P}\left(N = i\right)}{\sum_{j=1}^b \mathbb{P}\left(N = j\right)}. \label{eq:ProbsConv}
\end{equation*}
We denote the PH-distribution arising this way $F^{(l,m,b)}(y,n)$. Using the representation for the conditional representation in Equation  \ref{eq:DensenessPHCondRep}, and letting
 $p_{n} = \mathbb{P}\left(N = n\right)$, we have
\begin{align*}
    \abs{F^{(l,m,b)}(y,n)-F(y,n)} &= \abs{\tilde{F}^{(l,m)}_{Y\mid N = n}(y)\tilde{p}^{(b)}_{n}-F_{Y\mid N = n} p_{n}} \\
    &= \abs{\tilde{F}^{(l,m)}_{Y\mid N = n}(y)\tilde{p}^{(n)}_{n}- \tilde{F}^{(l,m)}_{Y\mid N =n}(y)p_n + \tilde{F}^{(l,m)}_{Y\mid N =n}(y)p_n -F_{Y\mid N = n}(y) p_{n} } \\
    &\le \abs{\tilde{F}^{(l,m)}_{Y\mid N = n}(y)\tilde{p}^{(n)}_{n}- \tilde{F}^{(l,m)}_{Y\mid N =n}(y)p_n} + \abs{\tilde{F}^{(l,m)}_{Y\mid N =n}(y)p_n -F_{Y\mid N = n}(y) p_{n}} \\
    &\le \abs{\tilde{p}^{(b)}_{n}- p_n} +  \abs{\tilde{F}^{(l,m)}_{Y\mid N =n}(y)-F_{Y\mid N = n}(y)}.
\end{align*}
We know $\abs{\tilde{F}^{(l,m)}_{Y\mid N =n}(y)-F_{Y\mid N = n}(y)} \to 0$ by Lemma \ref{lem:cond}. Considering Equation \eqref{eq:ProbsConv} it is also evident that $\abs{\tilde{p}^{(b)}_{n}- p_n} \to 0$ (notice that we may even take $b=l$ or $b=m$), thus establishing the desired convergence.
\end{proof}

\subsection{Augmenting the state space}  \label{sec:Augmenting}
It turns out that it is useful to find a phase-type representation for $Y$ such that the current state reflects the value of $N$, particularly when we move towards the estimation procedure for our joint model. This makes it possible to condition on the absorption state, and thereby also condition on the value of $N$, which simplifies calculations considerably. Consider the division of $\tilde{E}$ into $E^{+}$ and $E^{0}$. We can write $\boldsymbol{\alpha}$ and $\mathbf{T}$ according to the division
\begin{align*}
    \boldsymbol{\alpha} &= \begin{pmatrix} \boldsymbol{\alpha}^{+} & \boldsymbol{\alpha}^{0} \end{pmatrix} = \begin{pmatrix} \boldsymbol{\alpha}^{+} & \mathbf{0} \end{pmatrix}\\
    \mathbf{T} &= \begin{pmatrix} \mathbf{T}^{++} & \mathbf{T}^{+0} \\ \mathbf{T}^{0+} & \mathbf{T}^{00} \end{pmatrix},
\end{align*}
where the equality in the first line is due to the previous assumption on $\boldsymbol{\alpha}^{0}$. Similar to Equation \eqref{eq:TRV}, $\mathbf{T}^{++}$ gives the rates between states in $E^{+}$, and so on. If we now let $\text{diag}(\mathbf{T}^{++})$ be the diagonal of $\mathbf{T}^{++}$ we can split $\mathbf{T}^{++}$ into two matrices:
\begin{align*}
    \mathbf{T}^{++}_D &= \Delta\left( \text{diag}(\mathbf{T}^{++})\right) \\
    \mathbf{T}^{++}_A &= \mathbf{T}^{++}- \Delta\left( \text{diag}(\mathbf{T}^{++})\right),
\end{align*}
where $\Delta(\cdot)$ is the operator that turns a vector into a matrix with the elements of the vector in the diagonal and zeroes elsewhere. Naturally we have $\mathbf{T}^{++} = \mathbf{T}^{++}_D + \mathbf{T}^{++}_A$. Consider now the following matrix
\begin{align*}
    \tilde{\mathbf{T}}_1 &= \begin{pmatrix} \mathbf{T}^{++}_D & \mathbf{T}^{+0} & \mathbf{T}^{++}_A & \mathbf{0} \\
    \mathbf{0} & \mathbf{T}^{00} & \mathbf{T}^{0+} & \mathbf{0} \\
    \mathbf{0} & \mathbf{0} & \mathbf{T}^{++} & \mathbf{T}^{+0} \\
    \mathbf{0} & \mathbf{0} &  \mathbf{T}^{0+} & \mathbf{T}^{00}
    \end{pmatrix},
\end{align*}
and the following initial vector
\begin{equation*}
    \tilde{\boldsymbol{\alpha}}_1 = \begin{pmatrix}\boldsymbol{\alpha}^+ & \mathbf{0} & \mathbf{0} & \mathbf{0} \end{pmatrix}.
\end{equation*}
Clearly $(\tilde{\boldsymbol{\alpha}}_1, \tilde{\mathbf{T}}_1)$ is also a phase-type representation for $Y$, and of larger dimension than $(\boldsymbol{\alpha}, \mathbf{T})$. Let $\{\tilde{X}_t\}_{t \geq 0}$ denote the underlying Markov chain of this larger representation. If absorption happens from one of the two first blocks (that is, one of the first $p$ states), then $N=1$, and if absorption happens from one of the two last blocks then $N\geq 2$. By using this splitting of $\mathbf{T}^{++}$ $n+1$ times we obtain a representation with $p(n+1)$ states, where absorption in block $i$ means that $n = \lceil i/2 \rceil$. 
\begin{example}\rm
For illustrative purposes, $\tilde{\mathbf{T}}_4$ for $n=4$ is depicted below:
\begin{align*}
    \tilde{\mathbf{T}}_4 & = \begin{pmatrix} \mathbf{T}^{++}_D & \mathbf{T}^{+0} & \mathbf{T}^{++}_A & \mathbf{0} & \mathbf{0} &\mathbf{0} & \mathbf{0} & \mathbf{0} & \mathbf{0} &\mathbf{0}\\
    \mathbf{0} & \mathbf{T}^{00} & \mathbf{T}^{0+} & \mathbf{0} & \mathbf{0} &\mathbf{0} & \mathbf{0} & \mathbf{0} & \mathbf{0} &\mathbf{0} \\ 
    \mathbf{0} & \mathbf{0} & \mathbf{T}^{++}_D & \mathbf{T}^{+0} & \mathbf{T}^{++}_A & \mathbf{0} & \mathbf{0} &\mathbf{0} & \mathbf{0} & \mathbf{0} \\
    \mathbf{0} & \mathbf{0} & \mathbf{0} & \mathbf{T}^{00} & \mathbf{T}^{0+} & \mathbf{0} & \mathbf{0} &\mathbf{0} & \mathbf{0} & \mathbf{0}  \\
    \mathbf{0} & \mathbf{0}& \mathbf{0} &\mathbf{0} & \mathbf{T}^{++}_D & \mathbf{T}^{+0} & \mathbf{T}^{++}_A & \mathbf{0} & \mathbf{0} &\mathbf{0}\\
    \mathbf{0} & \mathbf{0} & \mathbf{0} &\mathbf{0} & \mathbf{0} & \mathbf{T}^{00} & \mathbf{T}^{0+} & \mathbf{0} & \mathbf{0} &\mathbf{0}  \\
    \mathbf{0} & \mathbf{0} & \mathbf{0} & \mathbf{0}& \mathbf{0} &\mathbf{0} & \mathbf{T}^{++}_D & \mathbf{T}^{+0} & \mathbf{T}^{++}_A & \mathbf{0}  \\
    \mathbf{0} & \mathbf{0} &\mathbf{0} & \mathbf{0} & \mathbf{0} &\mathbf{0} & \mathbf{0} & \mathbf{T}^{00} & \mathbf{T}^{0+} & \mathbf{0}  \\
    \mathbf{0} & \mathbf{0}& \mathbf{0} &\mathbf{0} & \mathbf{0} & \mathbf{0} & \mathbf{0} & \mathbf{0} & \mathbf{T}^{++} & \mathbf{T}^{+0} \\
    \mathbf{0} & \mathbf{0}& \mathbf{0} &\mathbf{0} & \mathbf{0} & \mathbf{0} & \mathbf{0} & \mathbf{0} & \mathbf{T}^{0+} & \mathbf{T}^{00} \\\end{pmatrix}.
\end{align*}
As we are going to condition on the absorbing state, the exit-rate vector is also of importance. Consider $\tilde{\mathbf{T}}_4$. Clearly the corresponding exit rate vector $\tilde{\mathbf{t}}_4 = -\mathbf{T}_4 \mathbf{e}$ has the following structure:
\begin{equation*}
    \tilde{\mathbf{t}}_4 = \begin{pmatrix} \mathbf{t}^{+} & \mathbf{t}^{0} & \mathbf{t}^{+} & \mathbf{t}^{0} & \mathbf{t}^{+} & \mathbf{t}^{0} & \mathbf{t}^{+} & \mathbf{t}^{0} &\mathbf{t}^{+} & \mathbf{t}^{0} \end{pmatrix}^{\prime}.
\end{equation*}
We now define $\mathbf{t}_4^{(n)}$ as the exit-rate vector, where all elements are set to zero except those in the two blocks that corresponds to $N = n$. In our example,
\begin{equation*}
    \tilde{\mathbf{t}}_4^{(4)} = \begin{pmatrix} \mathbf{0} & \mathbf{0} & \mathbf{0} & \mathbf{0} &\mathbf{0} & \mathbf{0} & \mathbf{t}^{+} & \mathbf{t}^{0} &\mathbf{0} & \mathbf{0} \end{pmatrix}^{\prime}.
\end{equation*}
Vectors of this type are crucial in what follows. 
\end{example}

In the sequel, we suppress the index on $\boldsymbol{\alpha}$, $\mathbf{T}$ and $\mathbf{t}$ when the value of $n$ is not ambiguous, to avoid overly tedious notation. 

\subsection{Joint Distribution} \label{sec:JointDist} 
This subsection uses the augmented representation of the previous subsection to derive the joint distribution in an elegant manner.
\begin{theorem}
The random vector $(Y,N)$, has density function
\begin{align}
    f(y,n) = \text{P}\left(Y \in [y, y+dy], N = n\right) = \tilde{\boldsymbol{\alpha}} \exp\left(\tilde{\mathbf{T}} y \right) \tilde{\mathbf{t}}^{(n)}, \label{eq:jointDens}
\end{align}
and joint distribution function
\begin{align}
    F(y,n) = \text{P}\left( Y <y, N < n\right) = \sum_{i=1}^{n-1} \tilde{\boldsymbol{\alpha}} \mathbf{I} \left(-\tilde{\mathbf{T}}\right)^{-1} \tilde{\mathbf{t}}^{(i)} - \tilde{\boldsymbol{\alpha}} \exp\left(\tilde{\mathbf{T}}y\right) \left(-\tilde{\mathbf{T}}\right)^{-1} \tilde{\mathbf{t}}^{(i)}. \label{eq:jointDist}
\end{align}
\end{theorem}
\begin{proof}
We first show Equation \eqref{eq:jointDens} using direct calculations on the augmented representation $\left(\tilde{\boldsymbol{\alpha}}, \tilde{\mathbf{T}} \right)$. Hereafter we integrate to obtain Equation \eqref{eq:jointDist}. To get Equation \eqref{eq:jointDens} we employ the same technique as in Equation \eqref{eq:PHDens} and condition on the starting state and the absorption state of the underlying  Markov jump process, $\{\tilde{X}_t\}_{t\geq 0}$, as follows. 
\begin{align*}
   & f\left(y,n\right) \\
    & = \mathbb{P}\left(Y \in [y, y+dy], N = n \right) \\
    &=\sum_{j =1}^{(n+1)p}\sum_{i = 1}^{(n+1)p} \mathbb{P}\left(Y \in [y,+dy], N =n \mid X_0 = i, X_{y^{-}} = j \right) \cdot \mathbb{P}\left(X_{y^{-}} = j \mid X_0 = i\right) \cdot \mathbb{P}\left(X_0  = i \right) \\
    &= \sum_{j=1}^{(n+1)p} \sum_{i=1}^{(n+1)p} \mathbf{1}_{j \in \{np+1, np+2, \ldots, (n+1)p\}} \cdot t_j \cdot \exp\left(\tilde{\mathbf{T}} y\right)_{ij} \alpha_i \\
    &= \sum_{j=np+1}^{(n+1)p} \sum_{i=1}^{\abs{E^+}} \alpha_i \exp\left(\tilde{\mathbf{T}} y\right)_{ij} t_j \\
    &= \tilde{\boldsymbol{\alpha}} \exp\left(\tilde{\mathbf{T}} y\right) \tilde{\mathbf{t}}^{(n)}.
\end{align*}
The joint distribution function is then obtained through the decomposition
\begin{equation*}
    \mathbb{P}\left[Y < y, N <n \right] = \sum_{i=1}^{n-1} \mathbb{P}\left[Y<y \mid N = n\right].
\end{equation*}
$\mathbb{P}\left(Y<y\mid N = n\right)$ is found by integrating Equation \eqref{eq:jointDens} with respect to y. We obtain
\begin{align*}
    F(y,n) &= \mathbb{P}\left(Y <y, N < n\right) \\
    &= \sum_{i=1}^{n-1} \int_{0}^{y}  \tilde{\boldsymbol{\alpha}} \exp\left(\tilde{\mathbf{T}} u\right) \tilde{\mathbf{t}}^{(i)} du\\
    &=\sum_{i=1}^{n-1} \left[\tilde{\boldsymbol{\alpha}}\exp(\tilde{\mathbf{T}} u) \tilde{\mathbf{T}}^{-1} \tilde{\mathbf{t}}^{(i)}\right]_0^{y} \\
    &= \sum_{i=1}^{n-1} \tilde{\boldsymbol{\alpha}}\exp\left(\tilde{\mathbf{T}} y\right) \tilde{\mathbf{T}}^{-1} \tilde{\mathbf{t}}^{(i)} - \tilde{\boldsymbol{\alpha}} \mathbf{I}\tilde{\mathbf{T}}^{-1} \tilde{\mathbf{t}}^{(i)} \\
    &= \sum_{i=1}^{n-1} \tilde{\boldsymbol{\alpha}} \mathbf{I} \left(-\tilde{\mathbf{T}}\right)^{-1} \tilde{\mathbf{t}}^{(i)} - \tilde{\boldsymbol{\alpha}} \exp\left(\tilde{\mathbf{T}}y\right) \left(-\tilde{\mathbf{T}}\right)^{-1} \tilde{\mathbf{t}}^{(i)}.
\end{align*}
\end{proof}

As we only consider absorption from block $i$ in each term of the sum, we generally get
\begin{equation*}
    \left(-\tilde{\mathbf{T}}_{n-1}\right)^{-1} \tilde{\mathbf{t}}_{n-1}^{(i)} \neq \mathbf{e},
\end{equation*}
and we can therefore not reduce the expression further, as for the respective univariate cases.

\begin{remark}\rm
If we wish to calculate the joint density function or distribution function for multiple values of $n$, then we can use one augmented representation instead of creating a new representation for each value of $n$. Let $m$ be the maximum value of $n$ that we want to calculate the density/distribution function for. We then calculate the joint density function as
\begin{equation*}
    f(y,n) = \tilde{\boldsymbol{\alpha}}_m \exp\left(\tilde{\mathbf{T}}_m y\right) \tilde{\mathbf{t}}_m^{(n)},
\end{equation*}
and the joint distribution function as
\begin{equation*}
    F(y,n) = \sum_{i=1}^{n-1} \tilde{\boldsymbol{\alpha}}_{m-1} \mathbf{I} \left(-\tilde{\mathbf{T}}_{m-1}\right)^{-1} \tilde{\mathbf{t}}_{m-1}^{(i)} - \tilde{\boldsymbol{\alpha}}_{m-1} \exp\left(\tilde{\mathbf{T}}_{m-1} y\right) \left(-\tilde{\mathbf{T}}_{m-1}\right)^{-1} \tilde{\mathbf{t}}_{m-1}^{(i)}.
\end{equation*}
\end{remark}

\subsection{Conditional Distributions} \label{sec:ConditionalDist}
The conditional distributions of the form $Y\mid N=n$ and $N\mid Y=y$ can be calculated by means of the marginal distributions and the joint distributions:
\begin{align*}
    \mathbb{P}\left(Y \in [y,y+dy] \mid N = n \right) &= \dfrac{\mathbb{P}\left[Y \in [y,y+dy], N = n\right]}{\mathbb{P}\left(N=n\right)} \\
    \mathbb{P}\left(N=n \mid Y \in [y,y+dy]\right) &= \dfrac{\mathbb{P}\left(Y \in [y,y+dy], N = n\right]}{\mathbb{P}\left[Y \in [y,y+dy) \right]}.
\end{align*}
The expression $\mathbb{P}\left(Y \in [y,y+dy] \mid N = n \right)$ can be written directly from the density of a phase-type distribution using an exit matrix instead of an exit vector. We consider the augmented generator $\tilde{\mathbf{T}}$ defined in Section \ref{sec:JointDist}. We now define $n_{\max}$ different absorbing states, such that absorption from block $i$ happens to absorbing state $i$ (state $p+i$). We define the exit-matrix $\mathbf{\tilde{T}}_0$ by
\begin{align*}
    \mathbf{\tilde{T}}_0 = \begin{pmatrix} \mathbf{t} & 0 & 0  & \cdots & 0 \\
    0 & \mathbf{t} & 0  & \cdots & 0 \\
    0 & 0 & \mathbf{t}  & \cdots & 0 \\
    \vdots & \vdots & \vdots & \ddots & \vdots \\
    0 & 0  & 0  & \cdots & \mathbf{t}\end{pmatrix}.
\end{align*}
To condition on $N=n$ now corresponds to condition on the absorption state being state $p+n$. That is, we can write
\begin{equation*}
    \mathbb{P}\left( Y \in [y, y + dy] \mid N = n\right) = \mathbb{P}\left(Y \in [y, y+dy] \mid X_{\infty} = p+i \right),
\end{equation*}
where $X_{\infty}$ is the absorbing state. From \cite{CondPH} we then have
\begin{align*}
    f_{Y\mid N = n}(y) &= f_{Y \mid X_{\infty}(y) = p+n} = \boldsymbol{\gamma}(n) \exp\left(-\mathbf{A}y\right) \mathbf{a} \\
    F_{Y\mid N = n}(y) &= F_{Y \mid X_{\infty}(y) = p+n} =  1-\boldsymbol{\gamma}(n) \exp\left(-\mathbf{A}y\right) \mathbf{e},
\end{align*}
with
\begin{align*}
    \gamma(n)_i &= \dfrac{\boldsymbol{\tilde{\alpha}}\left(-\tilde{\mathbf{T}}\right)^{-1} \mathbf{e} \tilde{\mathbf{t}}^{(n)}_i \tilde{\pi}_i}{v^*_n} \\
    \mathbf{A} &= \Delta^{-1}\left(\boldsymbol{\tilde{\pi}} \right) \tilde{\mathbf{T}}' \Delta\left(\boldsymbol{\tilde{\pi}}\right) \\
    \mathbf{a} &= -\mathbf{A}\mathbf{e}
\end{align*}
and
\begin{align*}
    \mathbf{v}^* &= \tilde{\boldsymbol{\alpha}} \left( -\tilde{\mathbf{T}}\right)^{-1} \tilde{\mathbf{T}}_0 \\
    \tilde{\boldsymbol{\pi}} &= \dfrac{\tilde{\boldsymbol{\alpha}}\left(-\tilde{\mathbf{T}}\right)^{-1}}{\tilde{\boldsymbol{\alpha}}\left(-\tilde{\mathbf{T}}\right)^{-1} \mathbf{e}}.
\end{align*}

\subsection{Moment Generating Function} \label{sec:MGF}
\begin{theorem}
The joint moment generating function (MGF) of $(Y,N)$ is given by
\begin{equation*}
    H(\theta_1,\theta_2) = \boldsymbol{\alpha}\left(-\Delta\left(\mathbf{c}\left(\theta_1,\theta_2)\right) \right) - \mathbf{S}_{\theta_2} \right)^{-1} \mathbf{s},
\end{equation*}
with 
\begin{align*}
    \mathbf{S}_{\theta_2}&= \mathbf{S} + \Delta(\mathbf{b}) \\
    c_i(\theta_1,\theta_2) &= \theta_1 \exp\left(-\theta_2 \mathbf{1}_{i \in E^{+}} \right),
\end{align*}
where $b_i = s_{ii}\left(\exp\left(-\theta_2 \mathbf{1}_{i \in E^{+}} \right) -1\right).$
\end{theorem}
\begin{proof}
We first condition on the starting state of $X_t$, that is we calculate
\begin{align*}
    H_i(\theta_1,\theta_2) = \mathbb{E}\left( \exp\left(\theta_1 Y + \theta_2 N\right)\mid X_0 = i\right).
\end{align*}
Note, that we can write $Y$ and $N$ as a sum of the contributions coming from the first state and the remaining part:
\begin{align*}
    Y &= Y_{i1} + Y_r \\
    N &= N_{i1} + N_r,
\end{align*}
where $Y_{i1} \sim \exp(-s_{ii})$ is the time spent in the first state, and $Y_r$ is the remaining part of $Y$. $N_{i1}$ is 1 if $i \in E^+$ and 0 if $i \in E^0$. $N_r$ is the remaining part of $N$.  We have that $Y_{i1}$, $(Y_r)$, $N_{i1}$ and $N_r$ are all independent, and therefore we can write
\begin{align*}
    H_i(\theta_1,\theta_2) &= \mathbb{E}\left(\exp\left(\theta_1\left(Y_{i1} + Y_r \right) + \theta_2\left(N_{i1} + N_r\right) \right)\right) \\
    &= \mathbb{E}\left(\exp\left(\theta_1 Y_{i1} \right)\mid X_0 = i\right) \cdot \mathbb{E}\left(\exp\left(\theta_2 N_{i1} \right) \mid X_0 = i\right)\\
    &\quad \times\mathbb{E}\left(\exp\left(\theta_1 Y_{r} \right) \mid X_0 = i\right) \mathbb{E}\left(\exp\left(\theta_2 N_{r} \right) \mid X_0 = i\right).
\end{align*}
The first two terms can be calculated as
\begin{align*}
    \mathbb{E}\left(Y_{i1} \mid X_0 =i \right) &= \dfrac{1}{1+\theta_1 s_{ii}^{-1}} \\
    \mathbb{E}\left(N_{i1} \mid X_0 = i \right) &= \exp(\theta_2 \mathbf{1}_{i \in E^+}).
\end{align*}
For the two remaining terms we note, that they can be written using $H_j(\theta_1,\theta_2)$ for $i \neq j$
\begin{align*}
    \mathbb{E}\left(\exp\left(\theta_1 Y_{r} \right) \mid X_0 = i\right) \mathbb{E}\left(\exp\left(\theta_2 N_{r} \right) \mid X_0 = i\right) = \dfrac{s_i}{-s_{ii}} + \sum_{i =1, i \neq j}^p \dfrac{s_{ij}}{-s_{ii}} H_j(\theta_1,\theta_2).
\end{align*}
Combining the above terms we obtain
\begin{align*}
    H_i(\theta_1,\theta_2) = \dfrac{1}{1+\theta_1 s_{ii}^{-1}}  \cdot  \exp(\theta_2 \mathbf{1}_{i \in E^+}) \left( \dfrac{s_i}{-s_{ii}} + \sum_{i =1, i \neq j}^p \dfrac{s_{ij}}{-s_{ii}} H_j(\theta_1,\theta_2)\right),
\end{align*}
which is equivalent to
\begin{align*}
    -\theta_1 \exp\left(-\theta_2 \mathbf{1}_{i \in E^+}\right) H_i(\theta_1,\theta_2) = s_i + \sum_{i =1, i \neq j}^p s_{ij} H_j(\theta_1,\theta_2) + s_{ii} H_i(\theta_1,\theta_2)\exp(-\theta_2 \mathbf{1}_{i \in E^+}).
\end{align*}
Using the notation specified in the statement of the theorem, this amounts to
\begin{align*}
    -\Delta(\mathbf{c}(\theta_1,\theta_2)) \mathbf{H}(\theta_1,\theta_2) = \mathbf{S}_{\theta_2}  \mathbf{H}(\theta_1,\theta_2) + \mathbf{s},
\end{align*}
    which leads to
\begin{align*}
    \mathbf{H}(\theta_1,\theta_2) = \left(-\Delta(\mathbf{c}(\theta_1,\theta_2)) - \mathbf{S}_{\theta_2}\right)^{-1} \mathbf{s}.
\end{align*}
As $\boldsymbol{\alpha}\mathbf{H}(\theta_1,\theta_2) = H(\theta_1,\theta_2)$ we finally arrive at
\begin{align*}
    H(\theta_1,\theta_2) = \boldsymbol{\alpha} \left(-\Delta(\mathbf{c}(\theta_1,\theta_2)) - \mathbf{S}_{\theta_2}\right)^{-1} \mathbf{s}.
\end{align*}
\end{proof}

\subsection{Estimation} \label{sec:Estimation}
The Expectation-Maximization (EM) algorithm (see \cite{mclachlan2007algorithm}) is a general-purpose maximum-likelihood estimation procedure which takes advantage of latent (or hidden) variables to speed up convergence. Both discrete and continuous phase-type distributions are usually estimated using an EM algorithm (although other notable exceptions may be mentioned, for instance, Bayesian approaches as in \cite{bladt2003estimation}), since the sample path of the underlying Markov process carries a range of latent variables. The EM algorithm for continuous phase-type distributions was developed in \cite{EM}, and then later extended to the right-censored case by \cite{olsson1996estimation}, to the multivariate case by \cite{Breuer}, and to the covariate-dependent case by \cite{bladt2022phase,bladt2022phase2}. The present algorithm is similar to the aforementioned references, with the key distinction being that we use the augmented representation $(\tilde{\boldsymbol{\alpha}}, \tilde{\mathbf{T}})$ instead of the original one, which complicates computations in some places and simplifies them in others.

By construction, $(Y,N)$ arises from a realization of the Markov jump process $\{X_{t}\}_{t \geq 0}$. This realization consists of the ordered sequence of states visited $\mathbf{i} =\{i_0,i_1,i_2, \ldots,i_m, p+1\} $, and the time spent in these states upon each visit $\mathbf{u} = \{u_0, u_1, u_2, \ldots, u_m\}$. We call $x = (\mathbf{i},\mathbf{u})$ the full sample path, since it provides enough information to reconstruct the process. Let $\boldsymbol{\theta} = \left(\boldsymbol{\alpha}, \mathbf{T} \right)$ be the parameters that have to be estimated. Notice that we do not attempt to estimate $p$ and $E^{+}$; instead these are hyper-parameters, which should be specified before the estimation procedure starts. Assume that the full sample path is known. Let 
\begin{align*}
    \lambda_i &= -t_{ii}, \quad  i \in \{1,2,\ldots p\},\\
    p_{ij} &= -\dfrac{t_{ij}}{t_{ii}},\quad  i,j \in \{1,2, \ldots, p\}, i \neq j, \\
    p_i &= -\dfrac{t_i}{t_{ii}} = -\dfrac{\sum_{k=1, k\neq i}^{p} t_{ik}}{t_{ii}}, \quad  i \in \{1,2,\ldots, p\}. 
\end{align*}
We can then write the likelihood of $\boldsymbol{\theta}$ given the full sample path as 
\begin{align*}
   \alpha_{i_0} \lambda_{i_0} \exp\left(\lambda_i u_0 \right) p_{i_0,i_1} \lambda_{i_1} \exp\left(-\lambda{i_1} u_1 \right) \cdots p_{i_0,i_1} \\
    &\quad\times\lambda_{i_1} \exp\left(-\lambda{i_1} u_1 \right) \cdots  p_{i_{m-1},i_m} \lambda_{i_m} \exp\left(-\lambda{i_m} u_m \right) t_{i_m}.
\end{align*}

Using alternative notation, and since we have $\lambda_i \cdot p_{ij}  = t_{ij}$, we can  rewrite the likelihood for $k$ sample paths as 
\begin{align*}
    L\left(\boldsymbol{\theta},\mathbf{x}\right) = \prod_{i=1}^k L\left(\boldsymbol{\theta};x_i \right) = \prod_{i=1}^{p} \alpha_{i}^{B_i} \left(\prod_{i=1}^{p} \prod_{j=1}^{p} t_{ij}^{N_{ij}} \exp\left(-t_{ij} Z_i\right)\right) \prod_{i=1}^p t_i^{N_i} \exp\left(t_i Z_i\right),
\end{align*}
where $B_i$ is the number of processes starting in state $i$, $N_{ij}$ is the total number of jumps from state $i$ to state $j$, $Z_i$ is the total time spend in state $i$ and $N_i$ is the number of jumps from state $i$ to state $p+1$. Notice that in this expression, the complete sample paths do not appear; instead, we have four sets of random variables, $B_i$, $N_{ij}$, $N_i$ and $Z_i$ which fully specify the likelihood of the path. We call these the sufficient statistics, since their knowledge allows us to calculate the likelihood without knowing the full sample path. It is standard from exponential family theory (cf. \cite{barndorff1973exponential}) that given $B_i$, $N_{ij}$, $N_i$ and $Z_i$, the maximum-likelihood estimates of $\hat{\alpha}_i$, $\hat{t}_{ij}$ and $\hat{t}_i$ are
\begin{align}
\begin{split}
    \hat{\alpha}_i &= \dfrac{B_i}{k}, \quad   i \in \{1,2,\ldots, p\}, \\
    \hat{t}_{ij} &= \dfrac{N_{ij}}{Z_i}, \quad   i,j \in \{1,2,\ldots, p\}, i\neq j, \\
    \hat{t}_i &=  \dfrac{N_i}{Z_i}, \quad   j \in \{1,2,\ldots, p\},
    \end{split} \label{eq:EMPars}
\end{align}
and the estimates $\hat{t}_{ii}$ follow from the relation $-\hat{\mathbf{T}}\mathbf{e} = \hat{\mathbf{t}}$. However, as we only observe $(\mathbf{Y},\mathbf{N})$ and not directly $B_i$, $N_{ij}$, $N_i$ and $Z_i$ we employ the EM algorithm to still take advantage of the simple exponential family theory with explicit maximum likelihood estimators. To this end, we replace the sufficient statistics with conditional expectations before calculating the maximum likelihood estimates, and then iterate back and forth between the two calculations. This procedure is known to converge, but though local optima may be encountered.

\subsection*{E-step}

The tedious part of the algorithm is to calculate the conditional expectations, which we do below.
\begin{align*}
    \mathbb{E}\left(B_i \mid Y=y, N=n\right) &= \mathbb{E}\left(\mathbf{1}_{X_0 = i} \mid Y = y, N = n\right) \\
    &= \dfrac{\mathbb{P}\left( Y \in [y, y+dy], N=n \mid X_0 = i\right)\mathbb{P}\left(X_0 = i\right)}{\mathbb{P}\left( Y \in [y, y+dy], N=n\right)} \\
    &= \dfrac{\alpha_i \mathbf{e}_i \exp(\tilde{\mathbf{T}}y) \tilde{\mathbf{t}}^{(n)}}{\tilde{\boldsymbol{\alpha}} \exp(\tilde{\mathbf{T}}y) \tilde{\mathbf{t}}^{(n)}},
\end{align*}
\begin{align*}
    &\mathbb{E}\left(Z_i \mid Y=y, N=n\right)\\ &=\mathbb{E}\left( \int_{0}^y \mathbf{1}\{X_u = i \} du \mid Y =y, N = n\right) du \\
    &= \int_{0}^y \dfrac{\mathbb{P}\left(Y \in [y,y+dy],N=n\mid X_u = i \right) \cdot \mathbb{P}\left(X_u = i\right)}{\mathbb{P}\left(Y \in [y,y+dy],N=n\right)} du \\
    &= \int_{0}^y \sum_{k=1}^n \dfrac{\mathbb{P}\left(Y \in [y,y+dy], N = n \mid \tilde{X}_u = (k-1)p + i\right)\cdot \mathbb{P}\left(\tilde{X}_u = (k-1)p + i \right)}{\mathbb{P}\left(Y \in [y,y+dy], N = n\right)} du \\
    &= \sum_{k=1}^n \int_{0}^y \dfrac{\mathbf{e}_{(k-1)p+i}^{\top} \exp\left(\mathbf{\tilde{T}}(y-u)\right) \mathbf{\tilde{t}}^{(n)} \boldsymbol{\tilde{\alpha}} \exp\left(\mathbf{\tilde{T}u}\right) \mathbf{e}_{(k-1)p+i}}{\boldsymbol{\tilde{\alpha}} \exp(\tilde{\mathbf{T}}y) \tilde{\mathbf{t}}^{(n)}} du.
\end{align*}
To calculate $\mathbb{E}\left(N_{ij} \mid Y = y, N =n\right)$, we first calculate $\mathbb{E}\left(\tilde{N}_{ij}, Y= y, N = n\right)$, where $\tilde{N}_{ij}$ is the number of jumps between states in the augmented representation $\left(\tilde{\boldsymbol{\alpha}}, \tilde{\mathbf{T}}\right)$. We use this to calculate $\mathbb{E}\left(\tilde{N}_{ij} \mid Y = y, N = n\right)$, which again can be summed to get $\mathbb{E}\left(N_{ij} \mid Y = y, N =n\right)$.
\begin{align*}
    &\mathbb{E}\left( \tilde{N}_{ij}, Y = y , N = n\right)\\
    &= \sum_{l = 1}^{n_{\max}p} \mathbb{E}\left(\tilde{N}_{ij} \mathbf{1}\{\tilde{X}_{y^{-} = l}, Y \in [y, y + dy], N = n \}  \right) \\
    &= \sum_{l=1}^{n_{\max} p }  \mathbb{E}\left(\tilde{N}_{ij} \mid \tilde{X}_{y^{-}} = l  \right) \cdot \mathbb{P}\left(Y \in [y, y + dy] \mid \tilde{X}_{y^{-}} = l  \right) \cdot \mathbb{P}\left( N = n \mid \tilde{X}_{y^{-}} = l  \right)  \cdot \mathbb{P}\left(\tilde{X}_{y^{-}} = l \right) \\
    &= \sum_{l = 1}^{n_{\max} p}\mathbb{E}\left(\tilde{N}_{ij} \mid \tilde{X}_{y^{-}} = l  \right) t_l \mathbf{1}_{l \in \{(n-1)p+1,\ldots, np} \mathbb{P}\left(\tilde{X}_{y^{-}} = l \right) \\
    &= \sum_{l = (n-1)p +1 }^{n p}\mathbb{E}\left(\tilde{N}_{ij}, \mathbf{1}\{\tilde{X}_{y^-} =1 \}\right) \tilde{t}^{(n)}_l,
\end{align*}
from which
\begin{align*}
    \mathbb{E}\left(  \tilde{N}_{ij} \mid  Y = y , N = n\right) &= \dfrac{\mathbb{E}\left( \tilde{N}_{ij}, Y = y , N = n\right)}{\mathbb{P}\left( Y \in [y, y+dy], N=n\right)} \\
    &= \dfrac{\sum_{l = (n-1)p +1 }^{n p}\mathbb{E}\left(\tilde{N}_{ij}, \mathbf{1}\{\tilde{X}_{y^-} =1 \}\right) \tilde{t}^{(n)}_l}{\mathbb{P}\left( Y \in [y, y+dy], N=n\right)} \\
    &= \sum_{k = 1}^p \alpha_k\dfrac{\sum_{l = (n-1)p +1 }^{n p}\mathbb{E}\left(\tilde{N}_{ij}, \mathbf{1}\{\tilde{X}_{y^-} =1 \} \mid X_0 = k \right) \tilde{t}^{(n)}_l}{\mathbb{P}\left( Y \in [y, y+dy], N=n\right)} \\
    &=\sum_{k = 1}^p \alpha_k\dfrac{\sum_{l = (n-1)p +1 }^{n p} \left( t_{ij} \mathbf{e}_{j}^{\top} \int_0^y \exp\left(\tilde{\mathbf{T}}(y-u)\right) \mathbf{e}_l \mathbf{e}_k^{\top} \exp\left( \tilde{\mathbf{T}}u\right)du \mathbf{e}_i\right) \tilde{t}^{(n)}_l}{\boldsymbol{\tilde{\alpha}} \exp(\tilde{\mathbf{T}}y) \tilde{\mathbf{t}}^{(n)}}\\
    &= \dfrac{ \left( t_{ij} \mathbf{e}_{j}^{\top} \int_0^y \exp\left(\tilde{\mathbf{T}}(y-u)\right) \mathbf{\tilde{t}^{(n)}}\boldsymbol{\tilde{\alpha}} \exp\left( \tilde{\mathbf{T}}u\right)du \mathbf{e}_i\right)}{\boldsymbol{\tilde{\alpha}} \exp(\tilde{\mathbf{T}}y) \tilde{\mathbf{t}}^{(n)}}.
\end{align*}
Thus,
\begin{align*}
    \mathbb{E}\left(N_{ij} \mid Y  = y, N = n\right) &= \sum_{k = 1}^n \mathbb{E}\left(\tilde{N}_{(k-1)p+i,(k-1)p+j} \mid Y  = y, N = n\right)\\
    &\quad + \mathbb{E}\left(\tilde{N}_{(k-1)p+i,kp+j} \mid Y  = y, N = n\right),
\end{align*}
where one of the terms in the sum is going to be zero due to the structure of $\mathbf{\tilde{T}}$. At last we have
\begin{align*}
    &\mathbb{E}\left(N_i \mid Y= y, N = n\right) = \mathbb{P}\left( X_{y^-} = i \mid Y = y, N = n\right) \\
    &= \dfrac{\mathbb{P}\left(Y \in [y,y+dy], N = n \mid X_{y^-} = i \right) \mathbb{P}\left( X_{y^{-}} = i \right)}{\mathbb{P}\left(Y \in [y, y + dy], N = n \right)} \\
    &= \dfrac{ \sum_{l=1}^{n_{\max}}\mathbb{P}\left(Y \in [y,y+dy], N = n \mid \tilde{X}_{y^-} = (l-1)p+i \right) \mathbb{P}\left( \tilde{X}_{y^{-}} = (l-1)p + i \right)}{\mathbb{P}\left(Y \in [y, y + dy], N = n \right)} \\
     &= \dfrac{ \sum_{l=1}^{n_{\max}}\mathbb{P}\left(Y \in [y,y+dy] \mid \tilde{X}_{y^-} = (l-1)p+i \right) \mathbb{P}\left(N = n \mid \tilde{X}_{y^-} = (l-1)p+i \right)\mathbb{P}\left( \tilde{X}_{y^{-}} = (l-1)p + i \right)}{\mathbb{P}\left(Y \in [y, y + dy], N = n \right)} \\
     &= \dfrac{\sum_{l=1}^{n_{\max}} \tilde{\boldsymbol{\alpha}} 
     \exp(\tilde{\mathbf{T}}y) \mathbf{e}_{(l-1)+i} t_{(l-1)+i} \mathbf{1}_{l = n}}{\boldsymbol{\tilde{\alpha}} \exp(\tilde{\mathbf{T}}y) \tilde{\mathbf{t}}^{(n)}} \\
     &=\dfrac{\tilde{\boldsymbol{\alpha}} \exp(\tilde{\mathbf{T}}y) \mathbf{e}_{(n-1)+i} t_{(n-1)+i}}{\boldsymbol{\tilde{\alpha}} \exp(\tilde{\mathbf{T}}y) \tilde{\mathbf{t}}^{(n)}}.
\end{align*}
For the calculation of $\mathbb{E}\left(Z_i \mid Y = y N = n \right)$ and $\mathbb{E}\left(N_{ij} \mid Y = y, N = n \right)$ where an integral expression should be evaluated, we use the usual approach where a matrix exponential is used to evaluate the integral. We denote
\begin{equation*}
    \tilde{\mathbf{J}}(y,n;\tilde{\boldsymbol{\alpha}}, \tilde{\mathbf{T}}) = \int_0^y \exp\left(\tilde{\mathbf{T}}(y-u)\right) \mathbf{\tilde{t}^{(n)}}\boldsymbol{\tilde{\alpha}} \exp\left( \tilde{\mathbf{T}}u\right)du,
\end{equation*}
and evaluate it using the Van Loan identity (cf. \cite{van1978computing}):
\begin{align*}
    \exp\left(\begin{pmatrix} \tilde{\mathbf{T}} & \tilde{\mathbf{t}}^{(n)} \tilde{\boldsymbol{\alpha}} \\ \mathbf{0} & \tilde{\mathbf{T}}\end{pmatrix} y \right) &= \begin{pmatrix} \exp(\tilde{\mathbf{T}}y) & \tilde{\mathbf{J}}(y,n; \tilde{\boldsymbol{\alpha}}, \tilde{\mathbf{T}}) \\
    \mathbf{0} &\exp(\tilde{\mathbf{T}}y) \end{pmatrix}.
\end{align*}
Thus, we get
\begin{align*}
    \mathbb{E}\left(N_{ij} \mid Y = y, N = n \right) &= \dfrac{\sum_{k=1}^n \tilde{t}_{ij}\mathbf{J}_{(k-1)p + j,(k-1)p + i}(y,n; \tilde{\boldsymbol{\alpha}}, \tilde{\mathbf{T}}) + t_{ij} \tilde{\mathbf{J}}_{kp + j,(k-1)p + i}(y,n; \tilde{\boldsymbol{\alpha}}, \tilde{\mathbf{T}})}{\boldsymbol{\tilde{\alpha}} \exp(\tilde{\mathbf{T}}y) \tilde{\mathbf{t}}^{(n)}}, \\
    \mathbb{E}\left(Z_i \mid Y = y, N = n \right) &= \dfrac{\sum_{k=1}^n\tilde{\mathbf{J}}_{(k-1)p + i,(k-1)p + i}(y,n; \tilde{\boldsymbol{\alpha}}, \tilde{\mathbf{T}}) }{\boldsymbol{\tilde{\alpha}} \exp(\tilde{\mathbf{T}}y) \tilde{\mathbf{t}}^{(n)}}.
\end{align*}
For M observations we obtain the corresponding conditional expectations
\begin{align*}
    \mathbb{E}\left(B_i \mid \mathbf{Y}=\mathbf{y}, \mathbf{N} = \mathbf{n}\right) &= \sum_{m=1}^M \dfrac{\alpha_i \mathbf{e}_i \exp(\tilde{\mathbf{T}}y_m)\tilde{\mathbf{t}}^{(n_m)}}{\tilde{\boldsymbol{\alpha}} \exp(\tilde{\mathbf{T}}y_m) \tilde{\mathbf{t}}^{(n_m)}}, \\
    \mathbb{E}\left(Z_i \mid \mathbf{Y} = \mathbf{y}, \mathbf{N} = \mathbf{n}\right) &= \sum_{m = 1}^M \dfrac{\sum_{k=1}^{n_m}\tilde{\mathbf{J}}_{(k-1)p + i,(k-1)p + i}(y_m,n_m; \tilde{\boldsymbol{\alpha}}, \tilde{\mathbf{T}}) }{\boldsymbol{\tilde{\alpha}} \exp(\tilde{\mathbf{T}}y_m) \tilde{\mathbf{t}}^{(n_m)}}, \\
    \mathbb{E}\left(N_{ij} \mid \mathbf{Y} = \mathbf{y}, \mathbf{N} = \mathbf{n}\right) &= \sum_{m=1}^M\dfrac{\sum_{k=1}^{n_m} t_{ij}\tilde{\mathbf{J}}_{(k-1)p + j,(k-1)p + i}(y_m,n_m; \tilde{\boldsymbol{\alpha}}, \tilde{\mathbf{T}}) + t_{ij}\tilde{\mathbf{J}}_{kp + j,(k-1)p + i}(y_m,n_m; \tilde{\boldsymbol{\alpha}}, \tilde{\mathbf{T}})}{\boldsymbol{\tilde{\alpha}} \exp(\tilde{\mathbf{T}}y_m) \tilde{\mathbf{t}}^{(n_m)}}, \\
    \mathbb{E}\left(N_i \mid \mathbf{Y} = \mathbf{y}, \mathbf{N} = \mathbf{n} \right) &= \sum_{m = 1}^M  \dfrac{\tilde{\boldsymbol{\alpha}} \exp(\tilde{\mathbf{T}}y_m) \mathbf{e}_{(n_m-1)+i} t_{(n_m-1)+i}}{\boldsymbol{\tilde{\alpha}} \exp(\tilde{\mathbf{T}}y_m) \tilde{\mathbf{t}}^{(n_m)}}.
\end{align*}
\subsection*{M-step}
The maximum likelihood estimates of the parameters are simply given by
\begin{align*}
    \hat{\alpha_i} &= \dfrac{\mathbb{E}\left(B_i \mid \mathbf{Y} = \mathbf{y}, \mathbf{N} = \mathbf{n} \right)}{M} \\
    \hat{t}_{ij} &= \dfrac{\mathbb{E}\left(N_{ij}\mid \mathbf{Y} = \mathbf{y}, \mathbf{N} = \mathbf{n}\right)}{\mathbb{E}\left(Z_{ij} \mid \mathbf{Y} = \mathbf{y}, \mathbf{N} = \mathbf{n}\right)} \\
    t_i &= \dfrac{\mathbb{E}\left(N_{i}\mid \mathbf{Y} = \mathbf{y}, \mathbf{N} = \mathbf{n}\right)}{\mathbb{E}\left(Z_{ij} \mid \mathbf{Y} = \mathbf{y}, \mathbf{N} = \mathbf{n}\right)} \\
    t_{ii} &= -\sum_{j = 1, j\neq i}^n t_{ij} - t_{i}.
\end{align*}

%% file: InsuranceExample.tex
\section{Application: Insurance Data} \label{sec:Insurance}
A natural application of our model comes from the insurance sector. Namely, we are interested in modeling the number of claims and claim amounts for a given portfolio. A widespread assumption in this kind of data is to assume that each claim size is independent of the number of claims associated with the policy. If we let $Y_j$ be the random variable that denotes the claim amount of the $j$-th, claim and if let $N$ be the random variable that denotes the number of claims, this assumption allows us to calculate quantities such as the \textit{pure premium} very efficiently through
\begin{align*}
    \mathbb{E}\left(\sum_{j=1}^N Y_j\right) &=\mathbb{E}\left(N\right) \mathbb{E}\left(Y\right).
\end{align*}
Then two independent models are fitted: one to describe the number of claims, and one to describe the claim amounts. A lot of attention has been drawn towards the correct specification for each individual model, but very few models target to model both at once. Notable exceptions such as in \cite{Czado,Kramer,Frees} use copula approaches, see also \cite{Pinquet,Verschuren} for non-copula approaches.

Our model allows us to describe the claim amount and the number of claims with a joint, dependent distribution, and the underlying Markov process has the interpretation of a latent claim evolution until the full claim amount is realized. We let $\bar{Y}$ be the random variable that denotes mean claim amount of a policy, so that we aim to describe $(\bar{Y},N)$ with our joint model. This now leads to a different way of calculating the pure-premium, as we now have
\begin{equation*}
    \mathbb{E}\left(\sum_{j=1}^N Y_j\right) = \mathbb{E}\left(N \bar{Y}\right)
\end{equation*}
Importantly, it should be noted, that the time spend between two visits to a state in $E^+$ is not the same as the amount of a single claim.

\subsection{Swedish motorbike insurance}
\subsubsection{Data Description}
\begin{figure}[!htbp] 
\begin{subfigure}{0.9\linewidth}
\centering
\includegraphics[width=0.5\linewidth]{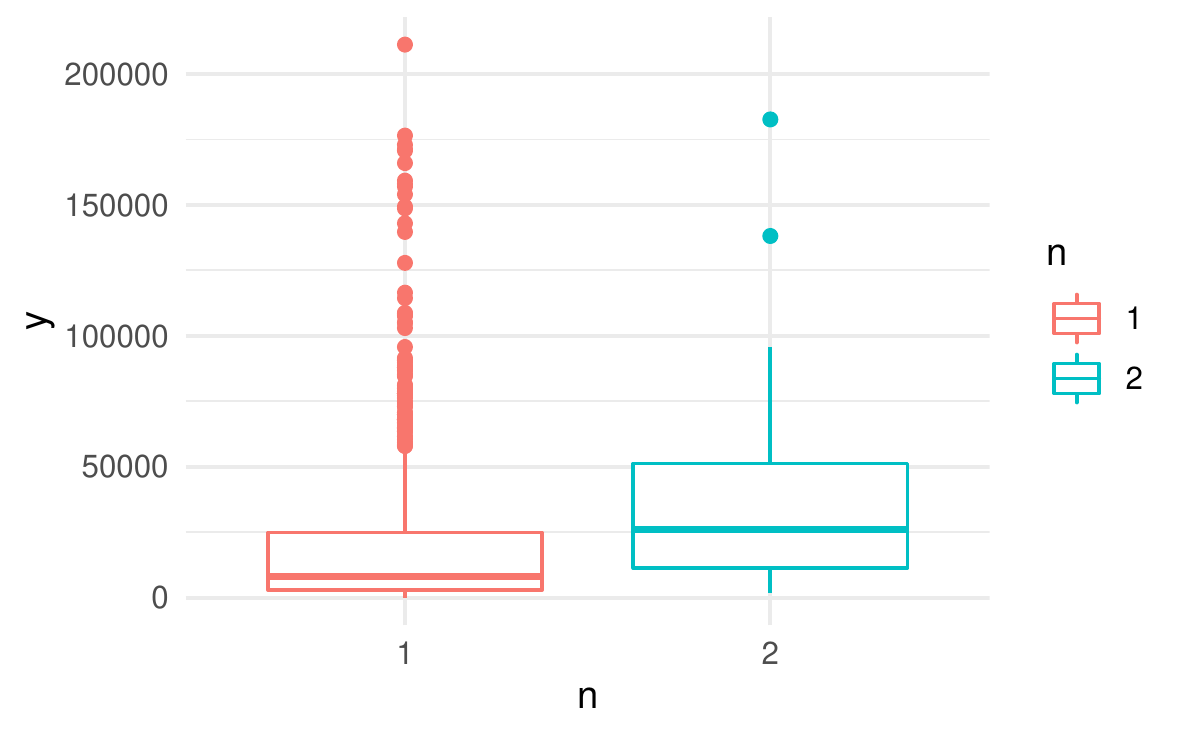}
\caption{Boxplots of $y$ for the different values of $n$} \label{fig:SweBoxY}
\end{subfigure}
\medskip
\begin{subfigure}{0.90\linewidth}
\centering
\includegraphics[width=0.5\linewidth]{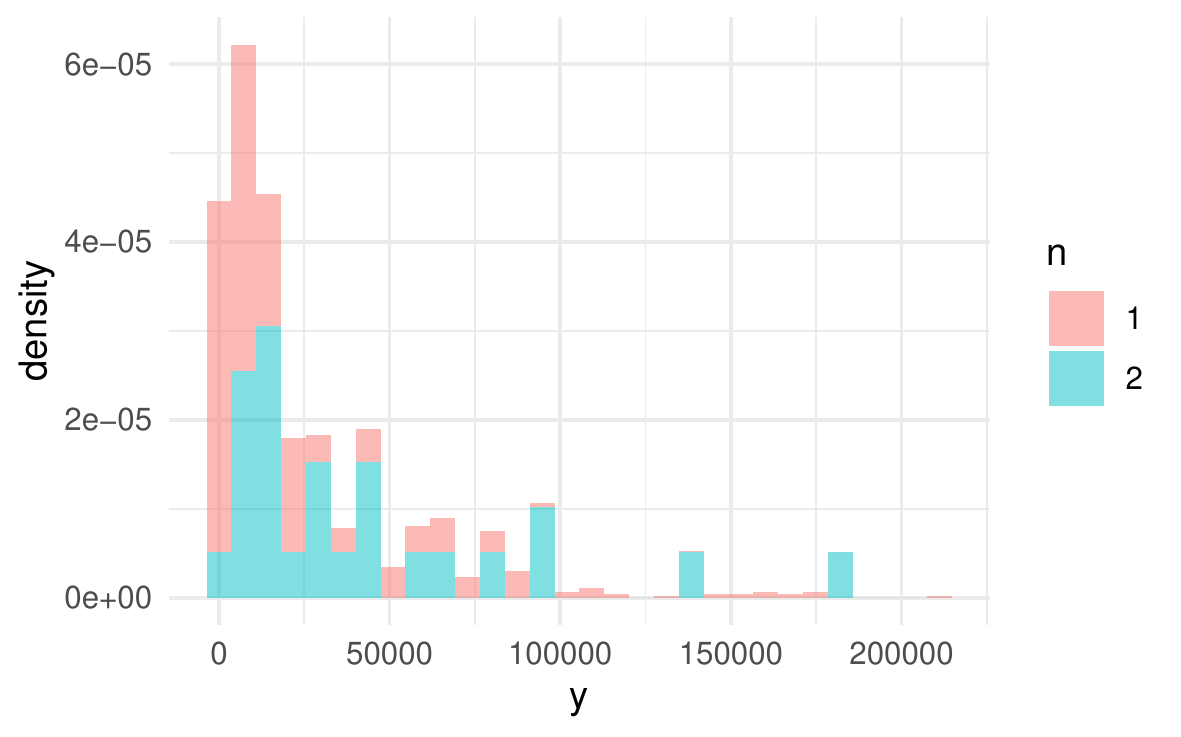}
\caption{Histograms of $\mathbf{y}$ for the different values of $n$} \label{fig:SweHistY}
\end{subfigure}
\medskip
\begin{subfigure}{0.9\linewidth}
\centering
\includegraphics[width=0.5\linewidth]{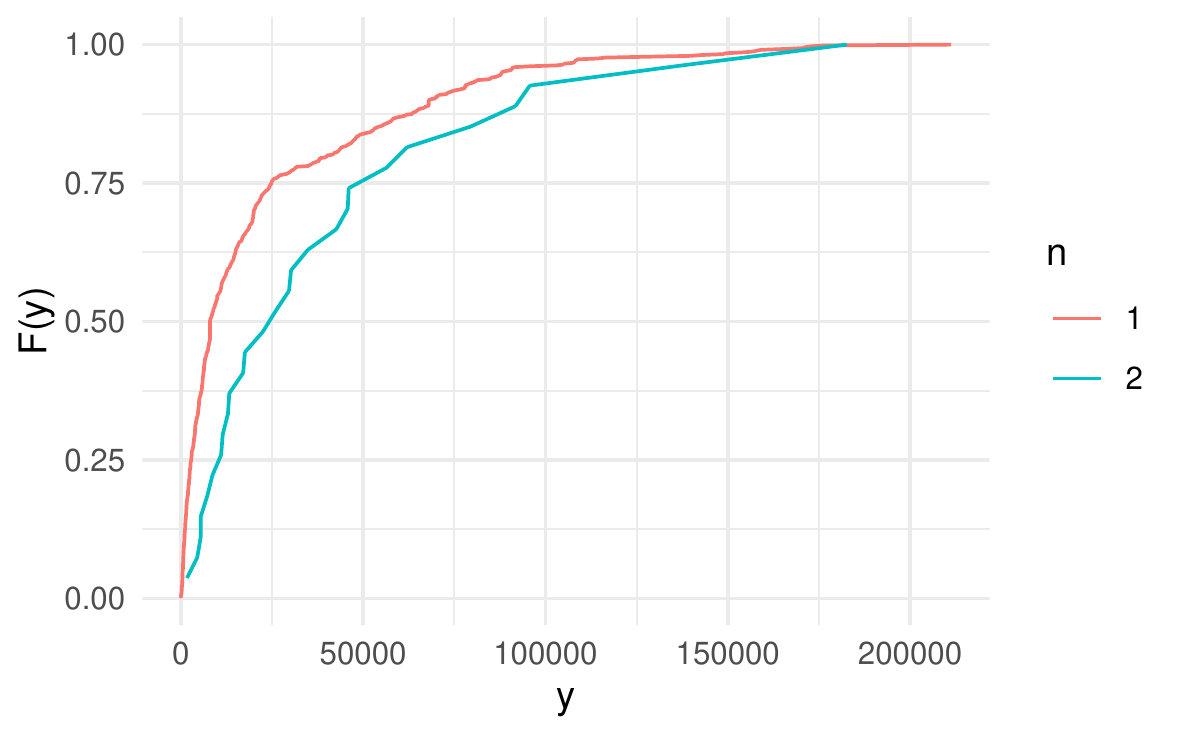}
\caption{The empirical CDF of $\mathbf{y}$ for the different }\label{fig:SweCDFY}
\end{subfigure}
\caption{Distributional graphs of $\mathbf{y}$ for the two different values of $n$.} \label{fig:SweY}
\end{figure}

The dataset $\texttt{swmotorcycle}$  from the \texttt{CASdatasets} \cite{CASdata} \texttt{R}-package contains claims for motorbike insurances. The dataset is of moderate size - there is a total of $666$ policies with claims. The manageable size makes it well-suited to investigate our method, since matrix analytic methods can result in heavy computational burden when the matrix dimensions grow. Let $n_i$ denote the number of claims for policy $i$ and let $y_i$ denote the mean-claim amount for policy $i$. Table \ref{tab:SweSummary} contains key information of the dataset based on the number of claims. 
\begin{table}[H]
    \centering
    \begin{tabular}{c|c c c c}
         $\#$ Claims & Mean of $\mathbf{y}$ & Median of $\mathbf{y}$ & Sd of $\mathbf{y}$  & $\mathbf{n}$\\ \hline
        1 & 23067 & 8050 & 33223 & 639 \\
        2 &  40769 & 25966 & 43925 & 27
    \end{tabular}
    \caption{Summary statistics for claims based on whether there was one or two claims to the policy. The mean and median of $\mathbf{y}$ are much higher for $n=2$ than for $n=1$ indicating that the independence assumption is not realistic.}
    \label{tab:SweSummary}
\end{table}
From table \ref{tab:SweSummary} it is clear that the mean claim amount is higher for $n_i=2$ than for $n_i=1$. We can also investigate the dependence between $(y,n)$ graphically. Figure \ref{fig:SweY} shows boxplots, histograms, and the empirical CDF's for the dataset. 

\subsubsection{Fitting the Model}
As mentioned earlier, we describe $(\mathbf{y},\mathbf{n})$ with our joint model. For numerical reasons, we subtract the minimum of the continuous value and add 1. This ensures that the data is positive and properly starts close to 0. We therefore consider
\begin{equation*}
    \tilde{y}_i = y_i - \min\left(\mathbf{y} \right) +1.
\end{equation*}
$\mathbf{n}$ only attains the values $1$ and $2$, and no standardisation is needed. It is not uncommon for insurance data to have single-digit claim counts, even for much larger portfolios. 

Before fitting the phase-type distribution one needs to select the matrix dimension, which means the number of phases, $p$, and the number of phases collecting rewards for the discrete distribution, $\abs{E^+}$. These hyperparameters are data-specific and therefore we experimented with different values in both cases to determine an appropriate size. We considered in $p \in \{2, \ldots, 10\}$ and $\abs{E^+} \in \{1,2,3\}$, and we only considered $\abs{E^+} < p$. For each combination of $(p, \abs{E^+})$ we fitted our joint distribution using the EM algorithm described in Section \ref{sec:Estimation}. 

To investigate the need for a joint model we also fitted an independent model consisting of a phase-type distribution of dimension $p$ to describe $\mathbf{y}$ and a discrete phase-type distribution of dimension $\abs{E^+}$ to describe $\mathbf{n}$. These models were fitted using the standard EM algorithms for continuous and discrete phase-type distributions, respectively. The joint likelihood of the independent model was then found by combining the likelihood of the two models for $\mathbf{y}$ and $\mathbf{n}$, and this quantity was then compared with the likelihood for the dependent joint model. For comparing models with different numbers of parameters one would typically use either AIC or BIC to determine which model is the preferred one. This approach does not work well with phase-type distributions, as adding one extra phase typically adds multiple extra parameters, something which is heavily penalized in AIC and even more so in BIC. We therefore instead look at likelihood evolution plots, to determine when the increase in the likelihood stagnates.

\begin{figure}[!htbp]
    \centering
    \includegraphics[width = 0.5 \linewidth]{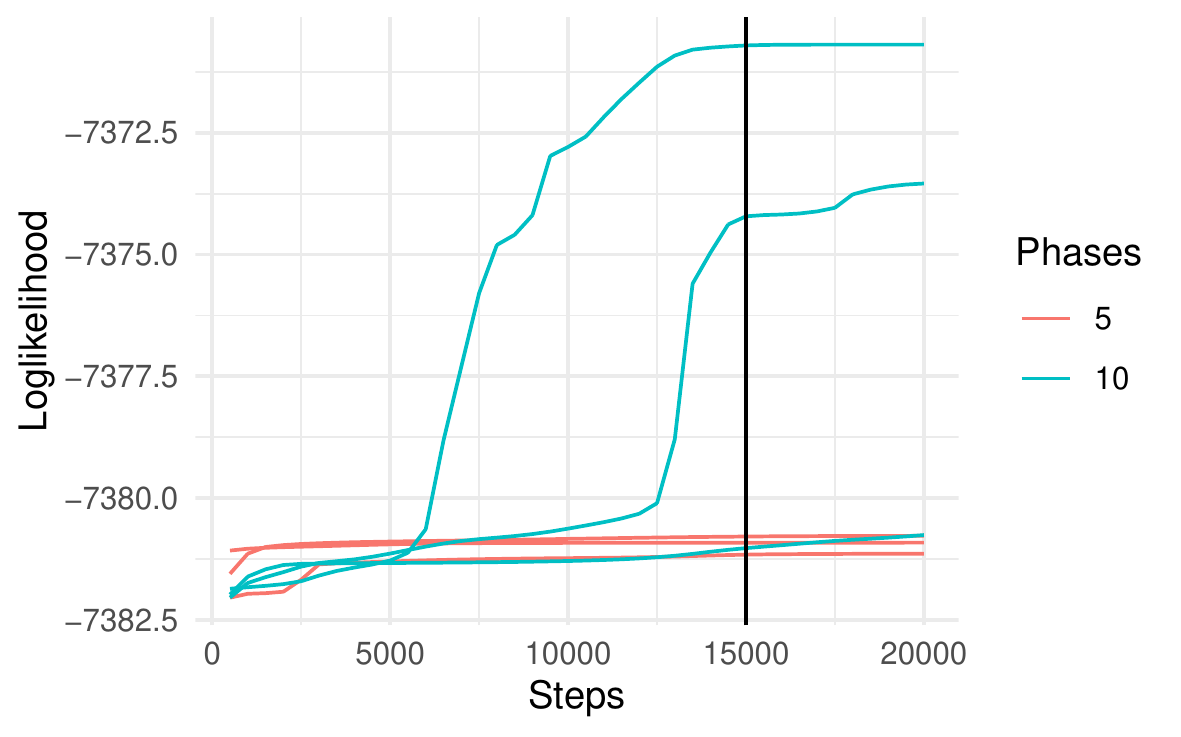}
    \caption{The log-likelihood as a function of the number of steps for a model with $p = 10$ (red) and $p = 5$ (blue). Three different start-guesses were used for each model. The figure illustrates that the likelihood of the final model is dependent on the start guess and that the small model converges faster than the large model. Finally, $15,000$ steps for the EM algorithm were chosen, as a compromise between the need for very precise convergence and computer resources. The black vertical line marks this. Furthermore, five different start guesses were considered for each model, and the one with the best likelihood was chosen.}
    \label{fig:llIterations}
\end{figure}

Through initial explorations of the EM algorithm, it was found that 
\begin{enumerate}
    \item The log-likelihood of the final fit depends on the start-guess
    \item The number of iterations needed for convergence depends on the model size and the start-guess
\end{enumerate}
Figure \ref{fig:llIterations} illustrates the previous statements. The figure shows the log-likelihood of the fit from the EM algorithm as a function of the number of steps for two different models - one with $p = 5$ and one with $p= 10$ - and three different start-guesses. In general the model with $p = 5$ converges faster than the model with $p = 10$, and the different start-guesses can lead to very different log-likelihoods. To overcome these challenges we decided to try five different random start-guesses for each model in the EM algorithm. The EM algorithm was then run with $15,000$ steps for each start-guess, and the fit with the best log-likelihood was then chosen. The number of steps was chosen by observing that most EM algorithms in Figure \ref{fig:llIterations} having converged already by then. 

\begin{figure}[!htbp]
    \centering
    \begin{subfigure}{0.6\textwidth}
    \includegraphics[width=\linewidth]{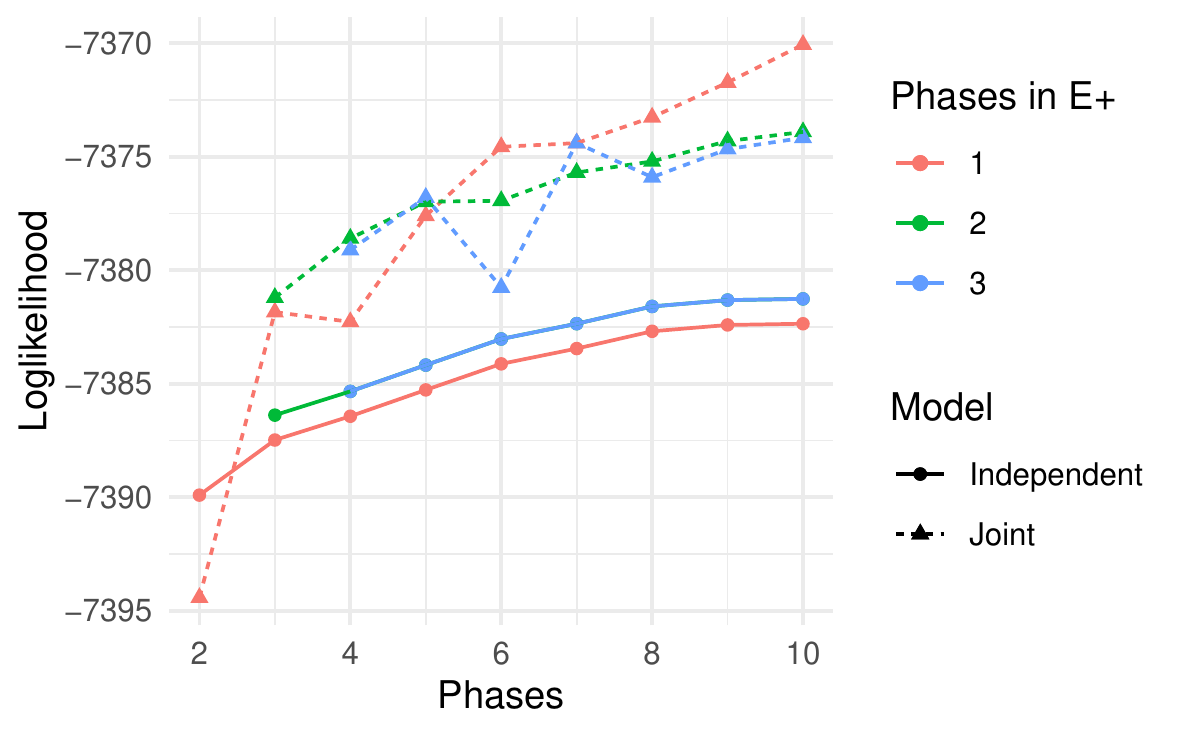}
    \caption{The loglikelihood against the model-sizes} \label{fig:ModelSize1}
    \end{subfigure}
    \medskip
    \begin{subfigure}{0.6\textwidth}
    \includegraphics[width=\linewidth]{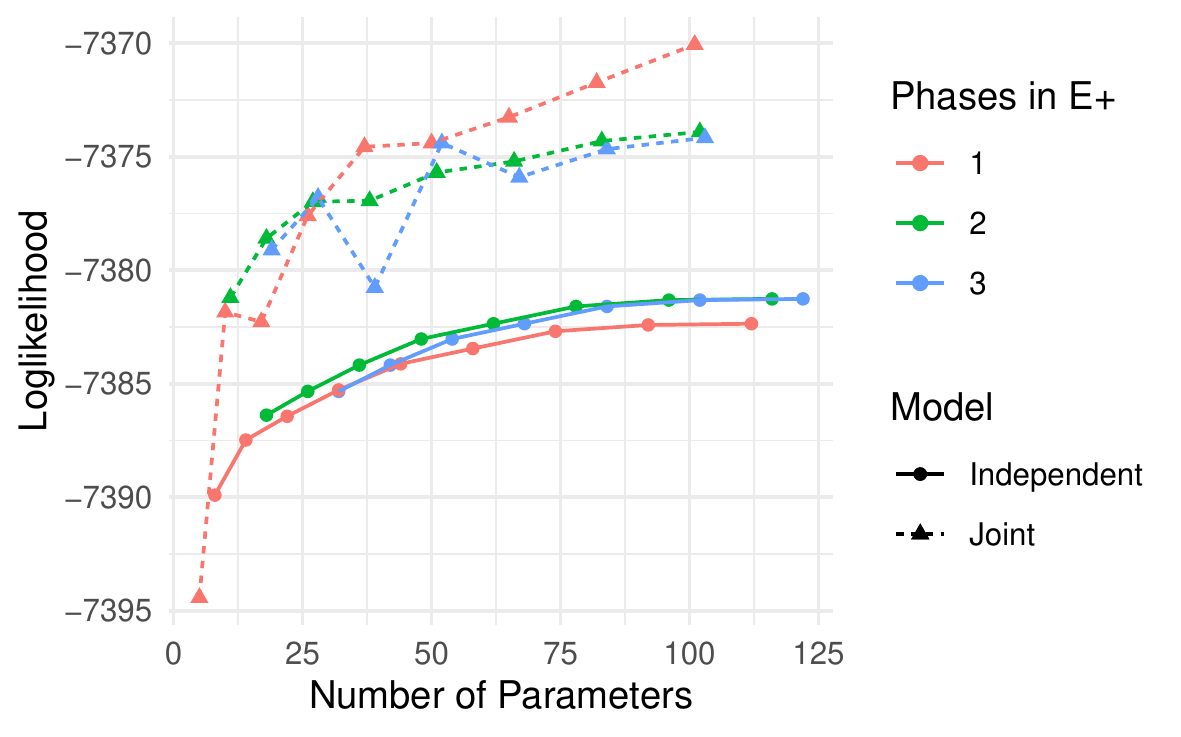}
    \caption{The loglikelihood against the number of parameters in the different models} \label{fig:ModelSize2}
    \end{subfigure}
    \caption{The log-likelihood for the independent model (solid lines) and the joint model (dashed lines) for different sizes. The top plot shows the log-likelihood against the number of phases, and the bottom plot shows the log-likelihood against the number of parameters. There is no difference between the independent model with $\abs{E^+} = 2$ and $\abs{E^+} = 3$ as $N$ only attains two values. Except for the model with $p = 2$ and $\abs{E^+} = 1$, the joint model fares better than the independent model in all cases. It also appears, unsurprisingly, that the estimation procedure for the joint model is more delicate than the usual EM algorithms used for the two components of the independent model ( though clearly much more competitive). For instance the likelihood for the joint model with $p = 6$ and $\abs{E^+} = 3$ which is clearly lower than expected and may be due to having reached a local minimum.}
    \label{fig:ModelSize}
\end{figure}

Figure \ref{fig:ModelSize} shows the result of fitting models of different sizes to the data. The dotted lines show the likelihood of the joint model and the solid lines show the likelihood of an independent model. The independent model consists of two phase-type distributions - one continuous and one discrete - and to ensure fairness they were also fitted with an EM algorithm using $15,000$ steps and with five different start-guesses. The continuous phase-type distribution was fit with the EM algorithm found in the \texttt{R}-package \texttt{matrixdist} (cf. \cite{matrixdist}), while the EM algorithm of the discrete phase-type distribution and the joint phase-type distribution was directly implemented by ourselves. One should note that the independent model has more parameters than the corresponding joint model as the number of parameters for the joint model is $p^2 + \abs{E^+}$, as we assume that the Markov chain starts in one of the states in $E^+$. The number of parameters in the independent models is $p^2 + p + \left(\abs{E^+}\right)^2 +  \abs{E^+}$. This asymmetry is considered below.

\subsubsection{Post Hoc Analysis}
Based on Figure \ref{fig:ModelSize} there is not one ideal joint model to work with. The likelihood increases with the number size of the model, but so does the number of parameters. We choose to work with the model with $p = 4$ and $\abs{E^+} = 2$. The best fit we obtained with these parameters had a log-likelihood of $-7378.599$, and lead to the following representation:
\begin{align}
\begin{split}
    \hat{\boldsymbol{\alpha}}_{4,2} &= \begin{pmatrix}
    0.94 & 0.06 & 0 & 0 
    \end{pmatrix}\\
    \hat{\mathbf{T}}_{4,2} & = \begin{pmatrix}
 -1.21\cdot 10^{-4} &   1.66 \cdot 10^{-6} & 0.00 &   3.63  \cdot 10^{-5} \\ 
0.00 &  -5.50\cdot 10^{-4}  & 0.00 &  0.00 \\
8.50 \cdot 10^{-7} &  4.03 \cdot 10^{-7} & -3.53 \cdot 10^{-5} & 0.00 \\
0.00 & 2.21 \cdot 10^{-6} & 3.31 \cdot 10^{-5} & -3.53 \cdot 10^{-4} 
    \end{pmatrix}, \end{split} \label{eq:fit42Rep}
\end{align}
with $\{1,2\} = E^{+}$. For comparison, the best fit for the independent model with $p = 4$ and $\abs{E^+} = 2$ had a log-likelihood of $-7385.338$, and many more parameters.

We can also calculate the mean time spent in each state using the Green matrix, $\mathbf{U} = (-\mathbf{T})^{-1}$, as follows:
\begin{equation*}
    \hat{\boldsymbol{\alpha}}_{4,2} \hat{\mathbf{U}}_{4,2} = \begin{pmatrix}  7817.415 & 172.4597 &  7633.909 &  8143.15 \end{pmatrix},
\end{equation*}
and we can calculate the mean number of visits to each state
\begin{equation*}
    \hat{\boldsymbol{\alpha}}_{4,2} \left( \mathbf{I}- \hat{\mathbf{Q}}_{4,2} \right)^{-1} = \begin{pmatrix} 0.95 &  0.10 &  0.27 &  0.29 \end{pmatrix},
\end{equation*}
where $\hat{\mathbf{Q}}_{4,2}$ is the transition matrix of the embedded Markov chain with elements as described in Equation \eqref{eq:EmbeddedTrans}. This, in particular, shows that no state is visited artificially, that is to generate a count without staying in the state for a non-negligible time. This feature is not always the case, as will be seen below.

\begin{figure}[!htbp]
\begin{subfigure}{0.48\textwidth}
\includegraphics[width=\linewidth]{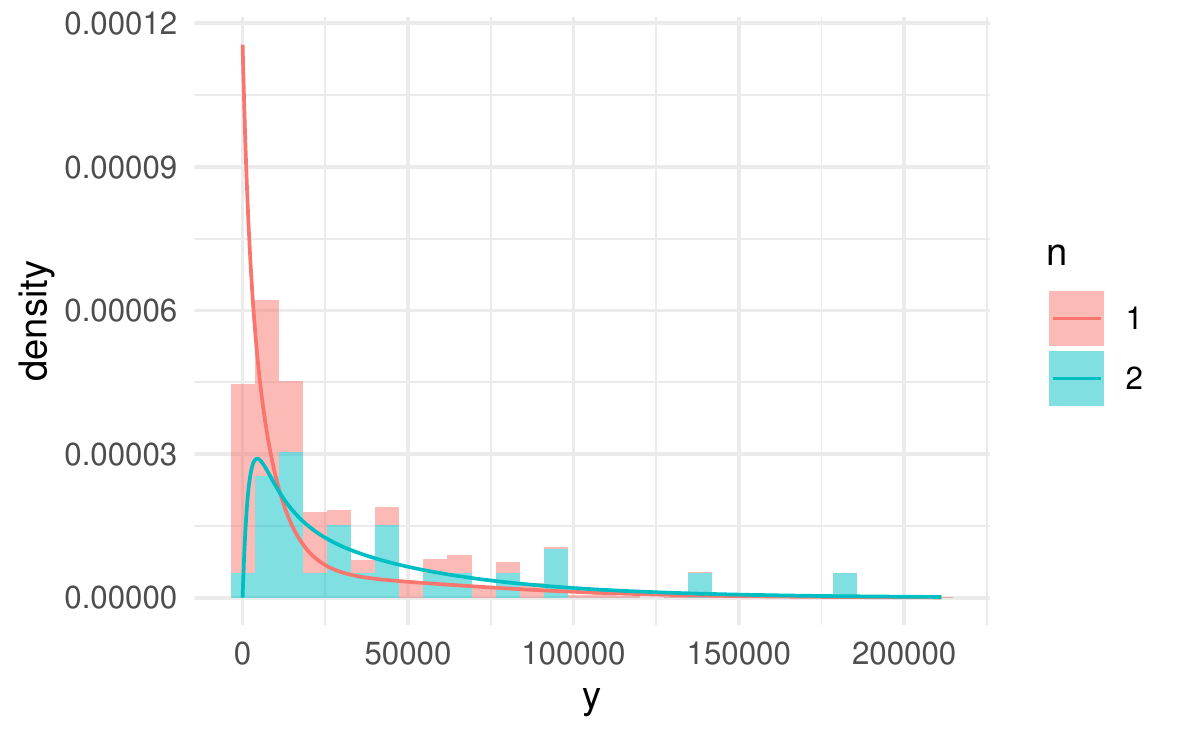}
\caption{Histograms of $\mathbf{y}\mid1$ and $\mathbf{y}\mid 2$ along with the conditional densities calculated from the estimated model\\ \\ \\ \\ \\}\label{fig:HistogramYFit}
\end{subfigure}\hspace*{\fill}
\begin{subfigure}{0.48\textwidth}
\includegraphics[width=\linewidth]{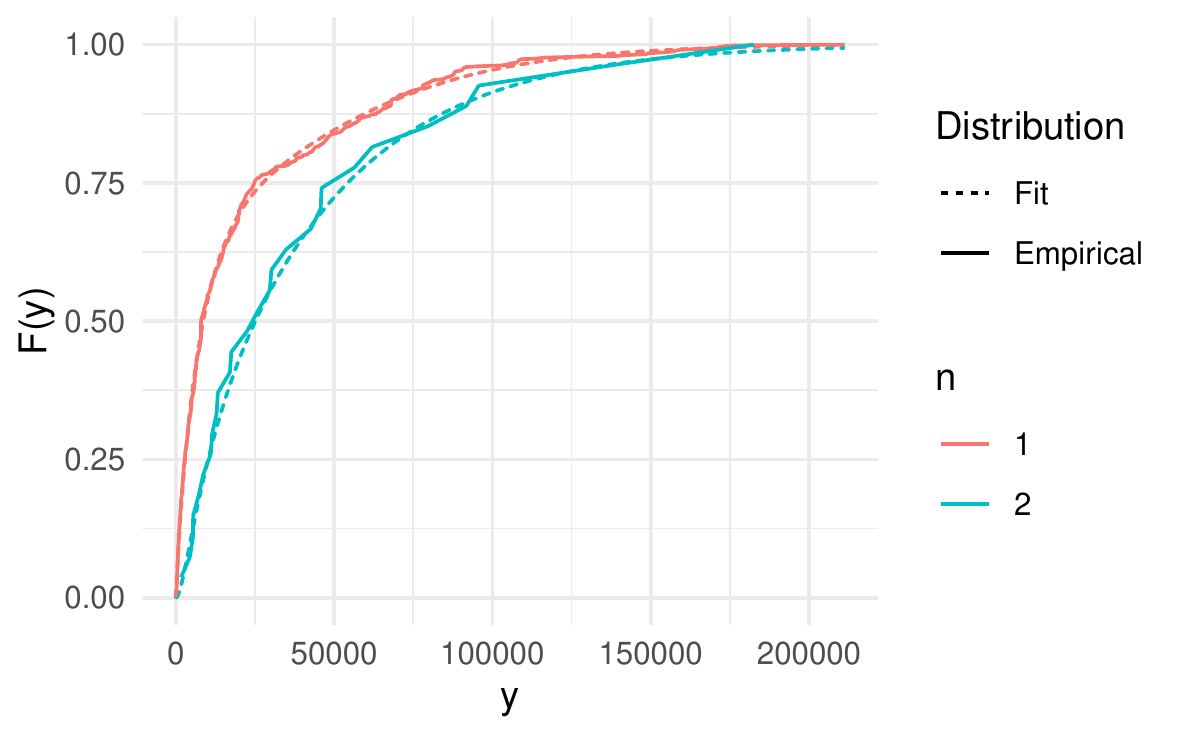}
\caption{The empirical CDF of $\mathbf{y}\mid 1$ and $\mathbf{y}\mid  2$ along with the CDF calculated from the theoretical distribution. For one claim the maximum distance between the two distribution functions is $D_1 = 0.032$, and for two claims the maximum distance between the two distribution functions is $D_2 = 0.055$. The distance is naturally higher for two claims because there are fewer observations.} \label{fig:CDFYFit}
\end{subfigure}
\caption{The conditional densities plotted over a histogram of the data, and the conditional CDF's plotted over the empirical CDF's.}
\label{fig:FitConditional}
\end{figure}

By the means of the formulas in Section \ref{sec:ConditionalDist} we can calculate the conditional distributions of $Y\mid N = 1$ and $Y \mid N = 2$, which are depicted in Figure \ref{fig:FitConditional}.

The joint model allows us to predict the total claim amount without the independence assumption. Instead we calculate the mixed moment using the moment generating function (see Section \ref{sec:MGF}). We obtain
\begin{equation*}
    \mathbb{E}\left(Y N\right) = 25396.084.
\end{equation*}
Under the independence assumption and with the corresponding independent model used to model $(Y,N)$ we would predict the total claim amount to be
\begin{equation*}
    \mathbb{E}\left(Y \cdot N \right) = \mathbb{E}\left(Y\right) \mathbb{E}\left( N\right) = 23769.21 \cdot  1.041 = 24732.84.
\end{equation*}
We can compare both these numbers to the actual claim amount, which is 
\begin{equation*}
    \dfrac{1}{m}\sum_{i=1}^m y_i n_i = 25421.4,
\end{equation*}
where $m$ is the number of observations. A standard $0.95$ confidence-interval is given by $[25307.45, 25535.35]$, which includes the prediction from the joint model but not the prediction from the independent model.

The introduction of covariates into our joint distribution is a subject of further research which would cast the model into a fully applicable pricing tool for actuaries.

\subsection{Discussion}
\subsubsection{Negative and positive dependence}
The dataset $\texttt{swmotorcycle}$ is a quite suitable dataset to describe with our joint model, as an initial exploratory analysis suggests that there is a difference between the distribution of $\mathbf{y}$ and $\mathbf{n}$ (see Figure \ref{fig:SweY} and Table \ref{tab:SweSummary}), and thus that the independence assumption does not hold. The dependence is interesting because $n=2$ is associated with higher values of $y$ than $n=1$, such that we have a positive correlation between $\mathbf{y}$ and $\mathbf{n}$. In a way this is the natural dependence in the joint model, as the states in $E^+$ increase both $n_i$ and $y_i$. It should naturally be possible to model negative correlations, as the MMPH$^{^\ast}$ class approximates all positive bivariate distributions with one continuous and one discrete component, though a satisfactory model might require more phases and thereby more parameters. 

To investigate this further, we change the dataset such that we consider $\tilde{n} = 3-n$ - that is if $n = 1$ then $\tilde{n} = 2$ and if $n = 2$ then $\tilde{n} =1$, to see how the model deals with this kind of negatively-correlated data. We do not run the entire experiment with different model sizes; instead we focus on the model with $p=4$ and $\abs{E^+} = 2$, which we also used on the original dataset. As before we try $5$ different start-guesses for the joint model, and the best log-likelihood obtained is $-7380.727$, which is lower than for the original dataset. The best log-likelihood for the independent model was $-7385.338$ like in the original dataset. Thus the joint model is still preferable, but the advantage decreases. Further gains require a full fledged analysis with respect to hyperparameter tuning. The best fit for the joint model on the switched dataset leads to the following representation:
\begin{align}
\begin{split}
    \hat{\mathbf{S}}_{4,2} &= \begin{pmatrix} 6.40 \cdot 10^{-4} & 0.9993\end{pmatrix}, \\
    \hat{\mathbf{S}}_{4,2} &= \begin{pmatrix} -1.36 \cdot 10^{-4} & 0.00 &  3.82 \cdot 10^{-5} &  0.00 \\
    4.82 & -5.02 & 7.31 \cdot 10^{-2}  &  1.27\cdot 10^{-1} \\
    0.00 & 0.00 &  -3.48\cdot 10^{-5} &  3.480\cdot 10^{-5} \\
     0.00 &  0.00 &   4.18 \cdot 10^{-7} &  -3.71 \cdot 10^{-5}\end{pmatrix},
     \end{split}\label{eq:fit42RepS}
\end{align}
with $E^+ = \{1,2\}$.

When comparing with the representation for the fit for the original dataset in Equation \eqref{eq:fit42RepS}, state $2$ stands out, which has a very large rate compared to the other states in both Equation \eqref{eq:fit42Rep} and in Equation \eqref{eq:fit42RepS}, meaning that the time spent in each visit is very small (the mean is $\approx 0.2$). As state 2 is in $E^+$, visits to this states thus increment $n$ without changing the size of $y$ significantly, which allows us to model the negative independence. This becomes even clearer when we consider the mean time spent in each state:
\begin{align*}
    \begin{pmatrix} 7052.167 &  0.20 &  8268.24 & 8445.41\end{pmatrix},
\end{align*}
and the mean number of visits to each state:
\begin{align*}
    \begin{pmatrix} 0.96 & 1.00 & 0.29 & 0.31 \end{pmatrix}.
\end{align*}
These vectors reveal that the chain always visits state $2$ once, which affects the value of $N$, but $Y$ remains relatively unchanged. In that way state $2$ becomes a state that is only targeting the modelling of $N$. This is also a reason as to why the log-likelihood gain is smaller in this case than with the original dataset: one phase is simply wasted on $N$ and therefore there are only three phases left to correctly model $Y$. In the representation found for the original dataset we also see that the mean time spent in state $2$ is small (though not as small as above), but this is due to fewer visits and not the rate. In short, one can say that negative dependence calls for states that are only contributing to $N$ and therefore more states are needed to get a high-performance fit.

\subsubsection{Size of discrete observations}
\begin{figure}[!htbp]
    \centering
    \includegraphics[width = 0.6 \linewidth]{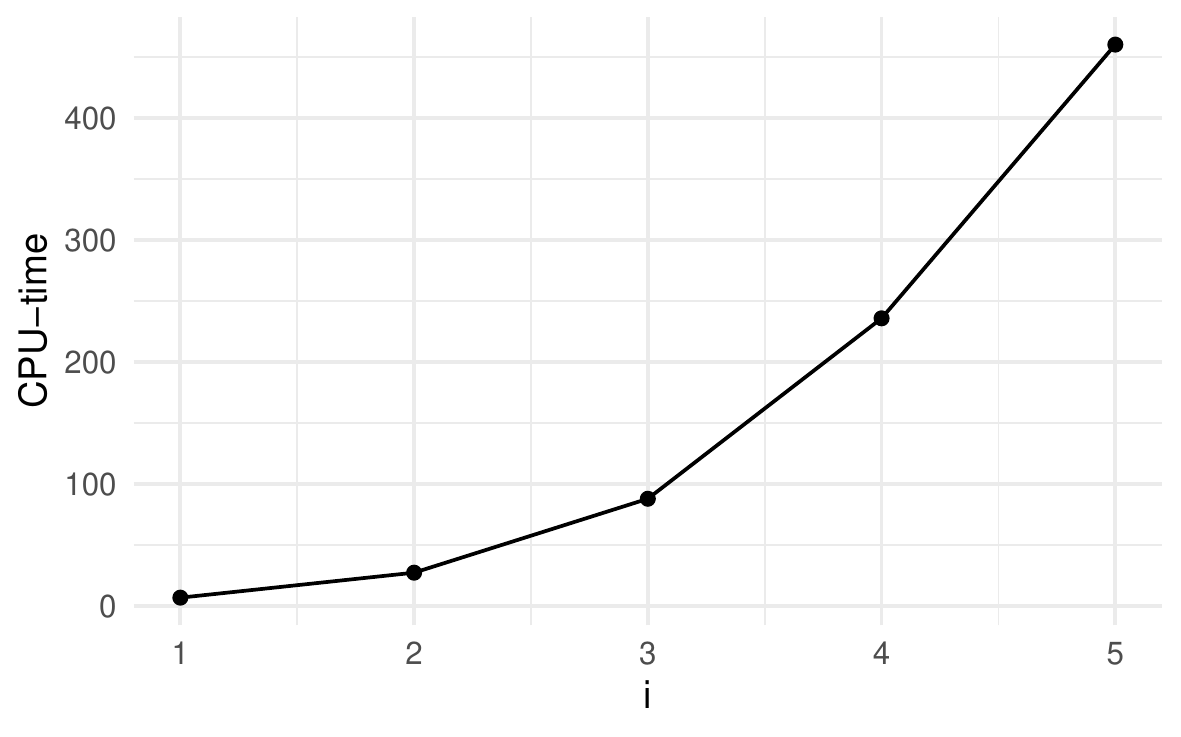}
    \caption{The CPU-time in seconds of the EM algorithm for fitting the joint model to simulated data. The simulated data was created by simulating $1000$ observations from a $\text{gamma}\left(4, \dfrac{1}{4} \right)$ distribution and using them as $\mathbf{y}$. All elements of $\mathbf{n}$ were set to $i$, and the figure shows the CPU-time as a function of $i$. Such an increase in computation burden is expected and in line with other phase-type EM algorithms, though still a possible practical drawback.}
    \label{fig:CPUtime}
\end{figure}

The EM algorithm for the joint model requires various matrix manipulations performed on $\tilde{\mathbf{T}}$ and therefore the time-consumption of the algorithm scales with the dimension of $\tilde{\mathbf{T}}$, which is $p \cdot (\max(\mathbf{n})+1)$. Thus the algorithm is be slower if the dataset which the EM algorithm is used on contains high values. For the \texttt{swmotorcycle} this is not problem as $\max(\mathbf{n}) = 2$, but for other kinds of data this might be and issue. Figure \ref{fig:CPUtime} illustrates this with simulated data. We simulated $1000$ observations from a $\text{gamma}\left(4,\dfrac{1}{4}\right)$ distribution and set all observations of $n = i$, $i = \{1,,\ldots,5\}$. The figure shows the CPU-time of $1000$ steps in the EM algorithm as a function of $i$.

%% file: Conclusion.tex
\section{Conclusion}\label{sec:con}
In this paper we propose a bivariate distribution on $\mathbb{R}_{+} \times \mathbb{N}$ with continuous and discrete phase-type distributed marginals. We proved that this class of distributions is dense in the class of distributions on $\mathbb{R}_{+} \times \mathbb{N}$. Hereafter we calculated the joint density and distribution function, conditional distributions, and the moment generating function. We also extended the EM algorithm for phase-type distributions such that it works in our bivariate setting. We applied our distribution to a dataset from the insurance industry, namely the average claim sizes and the number of claims of motorbike insurances. We compared our joint model to an independent model consisting of one continuous phase-type distribution and one discrete phase-type distribution and saw that for this particular dataset our joint model was substantially superior, shedding light on current actuarial practices. We used our framework to calculate the conditional distributions and saw how the model was able to correctly capture them, leading to a much improved estimate of the total loss. Finally, we investigated how the model fares if we consider a negatively correlated dataset. The conclusion was that the model performed worse than for the positively correlated case, but still better than the independent model even without any hyperparameter tuning.